\documentclass[prd,preprint,tightenlines,preprintnumbers,aps,eqsecnum,nofootinbib,superscriptaddress,floatfix]{revtex4-1}

\usepackage{amsmath}    % need for subequations
\usepackage{amssymb}
\usepackage{graphicx}   % need for figures
\usepackage{verbatim}   % useful for program listings
\usepackage{color}      % use if color is used in text
\usepackage{subfigure}  % use for side-by-side figures
\usepackage{longtable}
\usepackage{siunitx}
\usepackage[colorlinks=true,citecolor=blue,linkcolor=blue,urlcolor=blue]{hyperref}

\usepackage{bm}         % needed for better and more bold math symbols
\let\mathbf=\bm         % use the bm package's symbols instead

\usepackage{dsfont}

\definecolor{green}{rgb}{0.0,0.6,0.0}
\definecolor{blue}{rgb}{0.0,0.0,0.6}
\definecolor{gray}{rgb}{0.6,0.6,0.6}

\newcommand{\cut}[1]{}

\newcommand{\ahvp}{a_\mu^{\rm hvp}}
\makeatletter
\g@addto@macro\bfseries{\boldmath}
\makeatother

\begin{document}

\title{Rho resonance, timelike pion form factor, and implications for lattice studies of the hadronic vacuum polarisation}

\author{Felix~Erben}
\email{felix.erben@ed.ac.uk}
\affiliation{School of Physics and Astronomy, University of Edinburgh, Edinburgh EH9 3JZ, UK}
\affiliation{Helmholtz-Institut Mainz, 55099 Mainz, Germany}
\affiliation{Johannes Gutenberg-Universit\"at Mainz, 55099 Mainz, Germany}

\author{Jeremy~R.~Green}
\email{jeremy.green@desy.de}
\affiliation{NIC, Deutsches Elektronen-Synchrotron, D-15738 Zeuthen, Germany}

\author{Daniel Mohler}
\email{damohler@uni-mainz.de}
\affiliation{Helmholtz-Institut Mainz, 55099 Mainz, Germany}
\affiliation{Johannes Gutenberg-Universit\"at Mainz, 55099 Mainz, Germany}

\author{Hartmut~Wittig}
\email{hartmut.wittig@uni-mainz.de}
\affiliation{PRISMA$^+$ Cluster of Excellence and Institute for
  Nuclear Physics, Johannes Gutenberg-Universit\"at Mainz, 55099 Mainz, Germany}
\affiliation{Helmholtz-Institut Mainz, 55099 Mainz, Germany}

\preprint{MITP/19-062 \ \  DESY 19-165}
\noaffiliation

\date{\today}

\begin{abstract}
We study isospin-1 P-wave $\pi\pi$ scattering in lattice QCD with two
flavours of O($a$) improved Wilson fermions. For pion masses ranging
from $m_\pi=265$ MeV to $m_\pi=437$ MeV, we determine the energy
spectrum in the centre-of-mass frame and in three moving
frames. We obtain the scattering phase shifts using
L\"uscher's finite-volume quantisation condition. Fitting the
dependence of the phase shifts on the scattering momentum to a
Breit-Wigner form allows us to determine the resonance parameters
$m_\rho$ and $g_{\rho\pi\pi}$. By combining the scattering phase
shifts with the decay matrix element of the vector current, we calculate the
timelike pion form factor, $F_\pi$, and compare the results to the
Gounaris-Sakurai representation of the form factor in terms of the
resonance parameters. In addition, we fit our data for the form factor
to the functional form suggested by the Omn\`es representation, which
allows for the extraction of the charge radius of the pion. As a
further application, we discuss the long-distance behaviour of the
vector correlator, which is dominated by the two-pion channel.
We reconstruct the long-distance part in two ways: one based on the
finite-volume energies and matrix elements and the other based on
$F_\pi$. It is shown that this part can be accurately constrained
using the reconstructions, which has important consequences for lattice
calculations of the hadronic vacuum polarisation contribution to the
muon anomalous magnetic moment.
\end{abstract}


\maketitle

\section{Introduction}

The study of hadronic resonances in terms of the underlying theory of
QCD necessitates a non-perturbative treatment. Lattice QCD has emerged
as a versatile tool enabling \emph{ab-initio} determinations of many
hadronic properties \cite{Aoki:2019cca}. The $\rho$ meson, which is the simplest QCD resonance and decays almost exclusively into two pions
\cite{Tanabashi:2018oca}, is interesting for several reasons: It serves as a benchmark for the finite-volume formalism
pioneered by L\"uscher \cite{luscher1,luscher2,luscher_mf}, whose
practical implementation poses a number of challenging
tasks. Furthermore, the relevant correlation function have a rather
favourable noise-to-signal ratio compared to those for other
resonances, due to the $\rho$ being the lightest isovector resonance.

Beyond its role as a benchmark, the precision study of the $\rho$
resonance has a number of interesting applications. A good
understanding of the $\rho \rightarrow \pi\pi$ channel is a vital
component for any study of more complicated resonances, where the
$\rho$ is an intermediate decay channel. Thus, the $\rho$ has been
subject to many lattice QCD studies already~\cite{rpp-1, rpp-2, rpp-3,
  rpp-4, rpp-5, irrep-zetas, rpp-7, rpp-8, rpp-9,Fu:2016itp, Andersen:2018mau, Guo:2016zos, Alexandrou:2017mpi, Werner:2019hxc}. Secondly, using
the approach suggested by Meyer \cite{Meyer:timelike} (which is
closely related to work by Lellouch and L\"uscher
\cite{Lellouch:2000pv}), the pion form factor, $F_\pi$, can be
determined in lattice QCD in the timelike region. For first lattice
implementations of this method see~\cite{Feng:2014gba,
  Andersen:2018mau}.

An interesting and increasingly relevant application of lattice
calculations of $F_\pi$ arises in the context of \emph{ab initio}
determinations of the hadronic vacuum polarisation (HVP) contribution to the
muon's anomalous magnetic moment, $\ahvp$. The latter is accessible
via the (spatially summed) vector correlator $G(x_0)$~\cite{g-2:tmr-1,
  g-2:tmr-2, g-2:tmr-3}, which, at large Euclidean times $x_0$ is
dominated by the two-pion channel. Given sufficiently precise data for
$F_\pi$, one can accurately constrain the long-distance regime of
$G(x_0)$ which helps to significantly reduce both statistical
and systematic uncertainties in
lattice calculations of $\ahvp$ \cite{Meyer:2018til}.

The outline of this paper is as follows: In Section \ref{methodology}
we summarise the methods used for determining the isospin-1 scattering
phase shift and the timelike pion form factor from our lattice
calculations. Section \ref{rho} presents our results for the
scattering phase shift, while section \ref{timelike} contains the
results for the timelike pion form factor. Implications for the
calculation of the leading order HVP contribution to the muon
anomalous magnetic moment $a_\mu$ are discussed in Section
\ref{hvp}. Finally, Section \ref{outlook} summarises our results. Our
analysis supersedes previous preliminary results presented in
\cite{Erben:2016zue,Erben:2017hvr}.

\section{Methodology}
\label{methodology}
\subsection{Determination of the finite volume energy spectrum}

To study the $\rho$ resonance, we first need to extract a tower of low-lying
energy levels. The strategy we use is to build a matrix of correlation functions
using interpolating field operators with the quantum numbers of the $\rho$
meson. The lowest states of the spectrum can be extracted using the variational
method \cite{gevp1,gevp2,gevp3,gevp-window}. We start by forming a correlator matrix
\begin{align}
C_{ij}(t) = \langle O_i(t) O_j(0)^\dagger \rangle = \sum_{n=1}^\infty e^{-E_n t} \langle 0 | O_i | n \rangle \langle n | O_j^\dagger | 0 \rangle
\end{align}
from the correlators formed of interpolating operators $O_i(t)$ for the $\rho$ and $\pi\pi$ states in a given frame and then solve a generalised eigenvalue problem (GEVP) 
\begin{align}
C(t) \mathbf{v}(t,t_0) = \lambda(t,t_0) C(t_0) \mathbf{v}(t,t_0)
\end{align}
for this matrix. The $n^{th}$ eigenvalue $\lambda_n$ asymptotically
decays exponentially with the energy $E_n$ of the $n^{th}$ state. There are
different ways of choosing the parameter $t_0$ in the GEVP; one of them is to keep $t_0$ constant (the ``fixed-$t_0$ method'')
and another way is to use the ``window method'' \cite{gevp-window}, which keeps
the window width $t_w = t-t_0$ constant. For suitable choices, the latter
ensures that the leading excited state contamination to $\lambda_n(t,t_0)$ from the finite
correlator basis comes from $\Delta E_n=E_{N+1}-E_n$ \cite{gevp-window}, where $N$
is the size of the basis.

%The operator base needs to be particularly large for lattices with small pion
%masses, where the lower end of the spectrum is denser with many multi-hadron
%states present.
% To reliably determine the spectrum, if we want information on the $(n+1)^{th}$ energy level, we have to extract the $n$ lower lying states first.

For the operator basis we use \cite{ETMC:resonance_parameters} 

\begin{align}
\rho^0(\mathbf{P},t) &= \frac{1}{2 L^{3/2}} \sum_{\mathbf{x}} e^{-i \mathbf{P}\cdot \mathbf{x}} \left( \bar{u} \Gamma u - \bar{d} \Gamma d \right) (t) \, ,
\end{align}
where $\Gamma \in \{ \gamma_i, \gamma_0 \gamma_i\}$ and
\begin{align}
(\pi \pi)(\mathbf{p}_1,\mathbf{p}_2,t) =
\pi^+(\mathbf{p}_1,t)
\pi^-(\mathbf{p}_2,t)
-\pi^-(\mathbf{p}_1,t)
\pi^+(\mathbf{p}_2,t)
\, .
\end{align}
The momenta $\mathbf{p}_1$ and $\mathbf{p}_2$ of the single pions add
up to the frame momentum $\mathbf{P}$, i.e. $\mathbf{p}_1+\mathbf{p}_2=\mathbf{P}\equiv2\pi/L\mathbf{d}$. The single-pion interpolators are defined by
\begin{align}
\pi^+ (\mathbf{q},t) =&
\frac{1}{2 L^{3/2}} \sum_{\mathbf{x}} e^{-i \mathbf{q} \cdot \mathbf{x}}
\big( \bar{u} \gamma_5 d \big) (\mathbf{x},t)
\, , \\
\pi^- (\mathbf{q},t) =&
\frac{1}{2L^{3/2}} \sum_{\mathbf{x}} e^{-i \mathbf{q} \cdot \mathbf{x}}
\big( \bar{d}  \gamma_5 u \big) (\mathbf{x},t) \phantom{a} \, .
\end{align}

%\subsubsection{Correlator construction in moving frames irreducible representations}

In a finite hypercubic volume, the rotational symmetry $O(3)$ of the
continuum is reduced to that of a discrete subgroup. The operators are
therefore classified by the irreducible representations (irreps) of
the respective subgroup. The set of irreps depends on the momentum
frame used. In this work, we are using a centre-of-mass frame (CMF) as
well as moving frames with three different lattice momenta with a
maximum momentum of $\mathbf{P}^2=3 (2\pi/L)^2$,
i.e. $\mathbf{d}^2=3$. Correlation functions are computed for all such
moving frames that can be realised on a lattice of spatial size
$L$. Frames that share the same absolute momentum are averaged over.

In the rest frame, continuum operators $O^J$ with spin $J$ are subduced
\cite{irrep-1} into the irreps $\Lambda$ of the octahedral group via
\begin{align}
O^{[J]}_{\Lambda,\mu} = \sum_M \mathcal{S}_{\Lambda,\mu}^{J,M} O^{J,M}
\, ,
\end{align}
where $M$ are the magnetic quantum numbers of $J$,
$\mathcal{S}_{\Lambda,\mu}^{J,M}$ are the subduction coefficients and $\mu$ is
the row of the finite volume irrep $\Lambda$. The $J$ in
$O^{[J]}_{\Lambda,\lambda}$ is in brackets because, although it was produced
only from operators with spin $J$, the operator can now have an overlap with
all other spins which are contained in $\Lambda$ \cite{irrep-CG}.

In moving frames, there is a further reduction of symmetry, namely into the subgroup of the octahedral group that keeps $\mathbf{P}$ invariant \cite{irrep-CG}, which is referred to as the little group \cite{irrep-little-group}. To subduce continuum operators into the lattice irreps of the moving frame, we need helicity operators 
\begin{align}
O^{J,\lambda}(\mathbf{p}) = \sum_M \mathcal{D}^{(J)*}_{M\lambda}(R) O^{J,M}(\mathbf{p})
\, ,
\end{align}
where $\lambda$ is the helicity index, and
$\mathcal{D}^{(J)*}_{M\lambda}(R)$ is a Wigner-$\mathcal{D}$ matrix
\cite{wigner-d} for the transformation $R$ that rotates
$|\mathbf{p}|\hat{\mathbf{e}}_z$ into $\mathbf{p}$
\cite{irrep-helicity}. This allows a further subduction into little
group irreps $\Lambda$, forming a so-called subduced helicity operator 
\begin{align}
O^{J,P,|\lambda|}_{\Lambda,\mu}(\mathbf{p}) = \sum_{\hat{\lambda}=\pm\lambda} \mathcal{S}^{\tilde{\eta} \hat{\lambda}}_{\Lambda,\mu} O^{J,P,\hat{\lambda}}(\mathbf{p})
\, ,
\end{align}
where $P$ is the parity of $O^{J,P,\hat{\lambda}}(\mathbf{p}=0)$ and
$\tilde{\eta} = P(-1)^J$.  

We construct multiparticle operators from linear combinations of products of single-particle operators with definite momentum. A general $\pi\pi$ creation operator in an irrep $\Lambda$ can be written \cite{irrep-CG}
\begin{align}
(\pi\pi)^{[\mathbf{p}_1,\mathbf{p}_2]\dagger}_{\mathbf{P},\Lambda,\mu} = \sum_{\substack{\mathbf{p}_1 \in \{ \mathbf{p}_1 \}^* \\ \mathbf{p}_2 \in \{ \mathbf{p}_2 \}^* \\ \mathbf{p}_1 + \mathbf{p}_2 =\mathbf{P}}} 
= \mathcal{C} (\mathbf{P},\Lambda,\mu,\mathbf{p}_1,\mathbf{p}_2) \pi^\dagger(\mathbf{p}_1) \pi^\dagger(\mathbf{p}_2)
\, ,
\end{align}
where $\{ \mathbf{p}_{1,2} \}^*$ is the group orbit of $\mathbf{p}_{1,2}$,
i.e. the set of momenta that are equivalent under an allowed lattice rotation. $\mathcal{C}$ is a Clebsch-Gordan coefficient which couples the irreps $\Lambda_1$ and $\Lambda_2$ of the single-pion creation operators  $\pi^\dagger(\mathbf{p})$ with the irrep $\Lambda$ of the $(\pi\pi)^\dagger$ operator. These single-pion irreps are either the $A_1^-$ irrep of the cubic group for $\mathbf{p}=0$ or the $A_2$ irrep of the little group of $\mathbf{p}$ for $\mathbf{p} \neq 0$. The coefficients relevant for this work are listed in \cite{irrep-CG,irrep-CG-3E}.

In the isospin limit G-parity allows only contributions from odd partial waves \cite{irrep-rho}. Taking these reductions of symmetry into account, the relevant irreps of the $\rho \rightarrow \pi\pi$ channel, where $J^P = 1^-$ and where $l=1$ is the dominant contributing partial wave are listed in Table \ref{table:irreps-rho-pipi}.
\begin{table}[tb]
\centering
\begin{tabular}{|c|c|c|}
\hline 
$\mathbf{d}$ & $\Lambda(\mathbf{dim}(\Lambda))$\\ 
\hline 
$[000]$ & $T_1(3)$\\ 
\hline 
$[00n]$ & $A_1(1),E(2)$\\ 
\hline 
$[0nn]$ & $A_1(1),B_1(1),B_2(1)$\\
\hline 
$[nnn]$ & $A_1(1),E(2)$\\ 
\hline
\end{tabular}
\caption{Irreps in the various moving frames used in this study.}
\label{table:irreps-rho-pipi}
\end{table}

In addition to the correlator matrix $C(t)$, the calculation of the timelike
pion form factor in Section \ref{tff} also requires the matrix elements $\langle  J_\mu(\mathbf{x}=0,t) O_i^\dagger(0) \rangle$, both for the local (single-site) current, \begin{align}
J_\mu^l(x) = \frac 12 Z_V\bar\psi(x) \gamma_\mu\tau^3\psi(x)
\, ,
\label{eqn:pointlike-curr}
\end{align}
and the conserved (point-split) current 
\begin{align}
J_{\mu}^c(x+\tfrac{a}{2}\hat \mu) = \frac 14 (
\bar \psi (x+a\hat \mu) (1+\gamma_\mu) U_\mu(x)^\dagger \tau^3 \psi(x) 
- \bar \psi (x) (1-\gamma_\mu) U_\mu(x) \tau^3 \psi(x+a\hat \mu) 
)
\, ,
\label{eqn:pointsplit-curr}
\end{align}
where $\psi(x) = (u,\,d)^T$ and $\tau^3 = {\mathrm{diag}}(1,\,-1)$. In
analogy to the single-meson operators, the spatial components of the current
operators $J_{\mu}(x)$ are projected into the respective irreps $\Lambda$,
yielding $J^{\Lambda}(x)$. In what follows, the superscript $\Lambda$
will be omitted in all equations where the irreps are treated the same
way.

We extract the relevant information on the ground state and the first few
excited states from the eigenvectors $v_n(t)$ of the corresponding
eigenvalues $\lambda_n(t)$ determined via the solution of the GEVP
of $C(t)$. 

The former are used to define operators $X_n(t)$ that project on the
state with energy $E_n$:
\begin{align}
X_n(t) = v_n^\dagger O(t) = \sum_i v_{ni}^* O_i
\, .
\end{align}
The corresponding two-point function is defined as
\begin{align}
D_{nn}(t) = \langle  X_n(t) X^\dagger_n(0) \rangle = v_n^\dagger C(t) v_n
\, ,
\end{align}
which is the (approximate) projection of the correlation matrix $C_{ij}(t)$ onto the correlator corresponding to the $n^{th}$ state. We investigated the eigenvectors $v_n$ on each timeslice and have chosen to use the vectors from the earliest timeslice after which the absolute value of their components plateaued. At large times, remnant contributions from other states in $D_{nn}(t)$ are expected to be exponentially suppressed such that only the $n^{th}$ state survives:
\begin{align}
D_{nn}(t) \rightarrow |Z_n|^2 \exp(-E_n t)
\, .
\label{eqn:dnn}
\end{align}
$Z_n = \langle \Omega | X_n | n \rangle$ is an overlap factor with state $n$
of the optimised interpolating operator $X_n$. From an exponential fit to
$D_{nn}(t)$ we extract $|Z_n|$ for our further analysis. The operators $X_n$ are then used to form a two-point function with the current insertions
at the sink, 
\begin{align}
\langle J(t) X_n^\dagger(0) \rangle = \sum_i v_{ni} \langle J(t) O_i^\dagger(0) \rangle 
\, ,
\end{align}
which again has a large-time behaviour dominated just by one state:
\begin{align}
\langle J(t) X_n^\dagger(0) \rangle \rightarrow \langle \Omega | J(t) | n \rangle Z_n^* e^{-E_n t}
\, .
\end{align}
The timelike pion form factor requires the knowledge of the matrix element $\langle \Omega | J(t) | n \rangle$, which can either be extracted by fitting an exponential function to $D_{nn}(t)$ and $\langle J(t) X_n^\dagger(0) \rangle$ or by forming the ratio \cite{Andersen:2018mau}:
\begin{align}
R^{E_n}(t) &= \frac{\langle J(t) X_n^\dagger(0) \rangle}{\sqrt{D_{nn}(t)} e^{-\frac12 E_n t}}
\rightarrow \frac{Z_n^*}{|Z_n|} \langle \Omega | J(t) | n \rangle
\, .
\end{align}
We also computed two other ratios with the same asymptotic value proposed by
the authors of \cite{Andersen:2018mau}. Similar to that work, we find $R(t)$
produces the most precise plateaus of the three and is not reliant on the fit
to Equation \eqref{eqn:dnn} for the extraction of $Z_n$. Therefore we fit a constant to $|R^{E_n}(t)|^2 = |\langle \Omega | J(t) | n \rangle|^2$ to extract the plateau value, which we denote $|A_{n}|^2$.

\subsection{The distillation method}

The two-pion operators are non-trivial to compute, due to so-called
sink-to-sink quark lines which require all-to-all propagators to be
computed. To facilitate this task we are using the ``distillation'' \cite{dist1} and
stochastic Laplacian Heavyside (LapH) smearing \cite{dist2} methods. 

With distillation \cite{dist1}, a smearing matrix $\mathcal{S}_{xy}(t)$ is constructed in the following way: We start with the lattice spatial Laplacian,
\begin{align}
-\nabla^2(\mathbf{x},\mathbf{y},t) = 6 \delta_{\mathbf{x},\mathbf{y}} - \sum_{j=1}^3 (\tilde{U}_j(\mathbf{x},t) \delta_{\mathbf{x}+ \hat{j},\mathbf{y}} + \tilde{U}_j^\dagger(\mathbf{y},t) \delta_{\mathbf{x}- \hat{j},\mathbf{y}})
 \, ,
\end{align}
where the gauge fields $\tilde{U}$ have been smeared using
$3$ iterations of stout smearing \cite{Morningstar:2003gk} with smearing
parameter $0.2$. We then compute the lowest $N_\mathrm{ev}$ eigenmodes $v^{(k)}$, defined via
\begin{align}
\sum_{\mathbf{y}} -\nabla^2(\mathbf{x},\mathbf{y},t) v^{(k)}(\mathbf{y},t) = \lambda^{(k)}(t)v^{(k)}(\mathbf{x},t)
 \, .
\end{align}
The definition of the actual smearing matrix is
\begin{align}
\mathcal{S}_{xy}(t) = \sum_{k=1}^{N_\mathrm{ev}} v^{(k)}(\mathbf{x},t) v^{\dagger (k)}(\mathbf{y},t) \equiv V(t) V^\dagger(t)
 \, .
\end{align}
One main advantage of this approach is that this smearing matrix can be split and used to project propagators into the subspace spanned by the $N_\mathrm{ev}$ eigenvectors, a much smaller number than the $3 N_L^3$ colour fundamental fields on each timeslice which are naively needed to save a propagator.

Particularly on larger lattices, because the total computational cost scales with the cube of the
physical volume or higher for fixed smearing, distillation is often treated stochastically \cite{dist2}. In this approach
noise-partitioning (also referred to as \emph{dilution}) in the space spanned by the Laplacian eigenmodes
\cite{wilcox-dilution,dilution} is used to reduce the variance of the
stochastic estimator. With a suitable dilution scheme, using just one
  noise per quark line typically produces a statistical uncertainty due to the
  stochastic estimation of the quark propagation that is of the same size or
  smaller than the one from the Monte-Carlo path integral.

A quark line, i.e. a smeared-to-smeared propagator within stochastic Laplacian-Heavyside
(LapH)-smearing, can be computed via
\begin{align}
\mathcal{Q} & = \mathcal{S} D^{-1} \mathcal{S} =\sum_b E(\varphi^{[b]}(\rho) ( \varrho^{[b]}(\rho))^\dagger)
\, ,
\end{align}
with the LapH sink vectors $\varphi$ and the LapH source vectors $\varrho$,
\begin{align}
\varphi^{[b]}(\rho) & =  \mathcal{S} D^{-1} V P^{(b)} \rho \\
\varrho^{[b]}(\rho) & = V   P^{(b)} \rho
\, .
\end{align}
These in turn are constructed using the noise vectors $\rho$, and the dilution projectors $P^{(b)}$. One can use $\gamma_5$-hermiticity to reverse quark propagators, giving rise to alternative LapH source and sink vectors
\begin{align}
\bar{\varphi}^{[b]}(\rho) & = \gamma_5 \varphi^{[b]}(\rho)\,, \\
\bar{\varrho}^{[b]}(\rho) & = \gamma_5 \varrho^{[b]}(\rho)
\, ,
\end{align}
which give a different estimator for the quark line, $E(\bar\varrho \bar\varphi^\dagger)$. Meson functions can then be expressed via
\begin{align}
\mathcal{M}_\Gamma^{[b_1,b_2],(\rho_1,\rho_2)} (\mathbf{v}_1,\mathbf{w}_2;\mathbf{p},t)
= \Gamma_{\alpha \beta} \sum_{\mathbf{x}} e^{-i \mathbf{p} \cdot \mathbf{x}}
\mathbf{v}^{[b_1]}_{a \alpha, \mathbf{x}t}(\rho_1)^* 
\mathbf{w}^{[b_2]}_{a \beta, \mathbf{x}t}(\rho_2)
\, ,
\end{align}
where $\mathbf{v},\mathbf{w}$ are LapH source or sink vectors $\varrho,
\bar{\varrho}$ or $\varphi, \bar{\varphi}$. Single-meson correlation functions
are a product of two such meson functions, for example
\begin{align}
\langle\pi^+(t_f)\pi^-(t_0)\rangle&= \frac{1}{4L^3}\langle -
\mathcal{M}^{[b_1,b_2],(\rho_1,\rho_2)}_{\gamma_5} (\bar{\varphi_i},\varphi_j;t_F)
\mathcal{M}^{[b_1,b_2],(\rho_1,\rho_2)}_{\gamma_5} (\bar{\varrho_i},\varrho_j;t_0)^*  \rangle_{U,\rho}
\, ,
\end{align}
which uses the Einstein summation convention for the dilution indices
$b_1,b_2$. As our correlation functions can contain two-pion operators at
  both the source and the sink, we evaluate expressions with products of up to
  four meson functions.

Correlation functions with a vector current at the sink require a propagator that is not smeared at the sink. This can be computed via~\cite{Mastropas:2014fsa, Andersen:2018mau}
\begin{equation}
  D^{-1} \mathcal{S} = \sum_b E( \phi^{[b]}(\rho) ( \varrho^{[b]}(\rho) )^\dagger ),
\end{equation}
where $\phi$ is a LapH unsmeared sink vector,
\begin{equation}
  \phi^{[b]}(\rho) = D^{-1} V P^{(b)} \rho,
\end{equation}
and likewise $\bar\phi^{[b]}(\rho)=\gamma_5 \phi^{[b]}(\rho)$ yields an estimator for $\mathcal{S}D^{-1}$. 

\subsection{Gauge field configurations and distillation schemes}

We use three gauge field ensembles with 2 dynamical mass-degenerate light flavours of nonperturbatively
improved Wilson quarks \cite{Sheikholeslami:1985ij,Jansen:1998mx} generated by
the Coordinated Lattice Simulations (CLS) consortium using the DDHMC algorithm
and software package \cite{Luscher:2005rx,Luscher:2007es}. The ensembles
were generated with $\beta=5.3$ corresponding to a lattice spacing of
$a=0.0658(7)(7)\mathrm{fm}$ and Table \ref{ensembles} lists key parameters of
these ensembles along with the number of configurations used in our study.

\begin{table}[htbp]
\centering
\begin{tabular}{ccccccccc}
\hline 
\hline 
 & $T/a$ & $L/a$ & $m_\pi$ [MeV]  & $\kappa$ & $m_\pi L$ & $N_\mathrm{conf}$ &
$N_\mathrm{meas}$ & $N_{\mathrm{ev}}$\\ 
\hline 
E5 &64 & 32 &  437 &  0.13625 & 4.7 & 500 & 2000 & 56\\ 
\hline 
F6 & 96 & 48 & 311 &  0.13635 & 5.0 & 300 & 900 & 192\\ 
\hline 
F7 & 96 & 48 & 265 &  0.13638 & 4.2 & 350 & 1050 & 192\\ 
\hline 
\hline 
\end{tabular} 
\caption{CLS $N_f=2$ ensembles used in this study. All share $\beta=5.3$ and
  $a=0.0658(7)(7)\mathrm{fm}$. $T$ and $L$ refer to the lattice extent in time
and space directions respectively. $N_\mathrm{conf}$ specifies the number of
gauge configurations used, while $N_\mathrm{meas}$ refers to the number of
source timeslices multiplied by the number of configurations. The number of Laplacian
eigenmodes used is denoted by $N_\mathrm{ev}$.}
\label{ensembles}
\end{table}

We use different dilution schemes for quark lines connected to the source
timeslice and for sink-to-sink quark lines. Lines connected to the source timeslice use full spin dilution
and full time dilution. Full Laplacian eigenvector dilution is used on E5, while
interlace-12 eigenvector dilution (\emph{LI12} in the notation of
\cite{dist2}) is used on F6 and F7.
The perambulators for sink-to-sink (sts) quark lines are calculated with full
spin dilution and interlace-8 time dilution (\emph{TI8}). On E5
sink-to-sink lines use \emph{LI8}, while \emph{LI12} is used on F6 and F7.

For the calculation of Laplacian eigenmodes, the PRIMME package \cite{PRIMME} is
used with a preconditioner built from Chebyshev polynomials
\cite{Neff:2001zr}. Our code uses the library QDP++ from USQCD \cite{qdp} and the deflated
SAP+GCR solver from the openQCD package \cite{openQCD}. For cross-checks of
the analysis the package TwoHadronsInBox \cite{Morningstar:2017spu} was used.

\section{The $\rho$ resonance}
\label{rho}

In this section the determination of the energy spectra, the
calculation of the phase shift from the energies, and the resulting resonance
mass $m_\rho$ and coupling $g_{\rho\pi\pi}$ are described.

\subsection{Energy spectra}

\begin{figure}[t!]
\centering
\leavevmode
%\includegraphics[width=0.45\textwidth]{F6_1_E2_window.pdf}
%\hfill
%\includegraphics[width=0.45\textwidth]{F6_2_A1_window.pdf}
\includegraphics[width=0.9\textwidth]{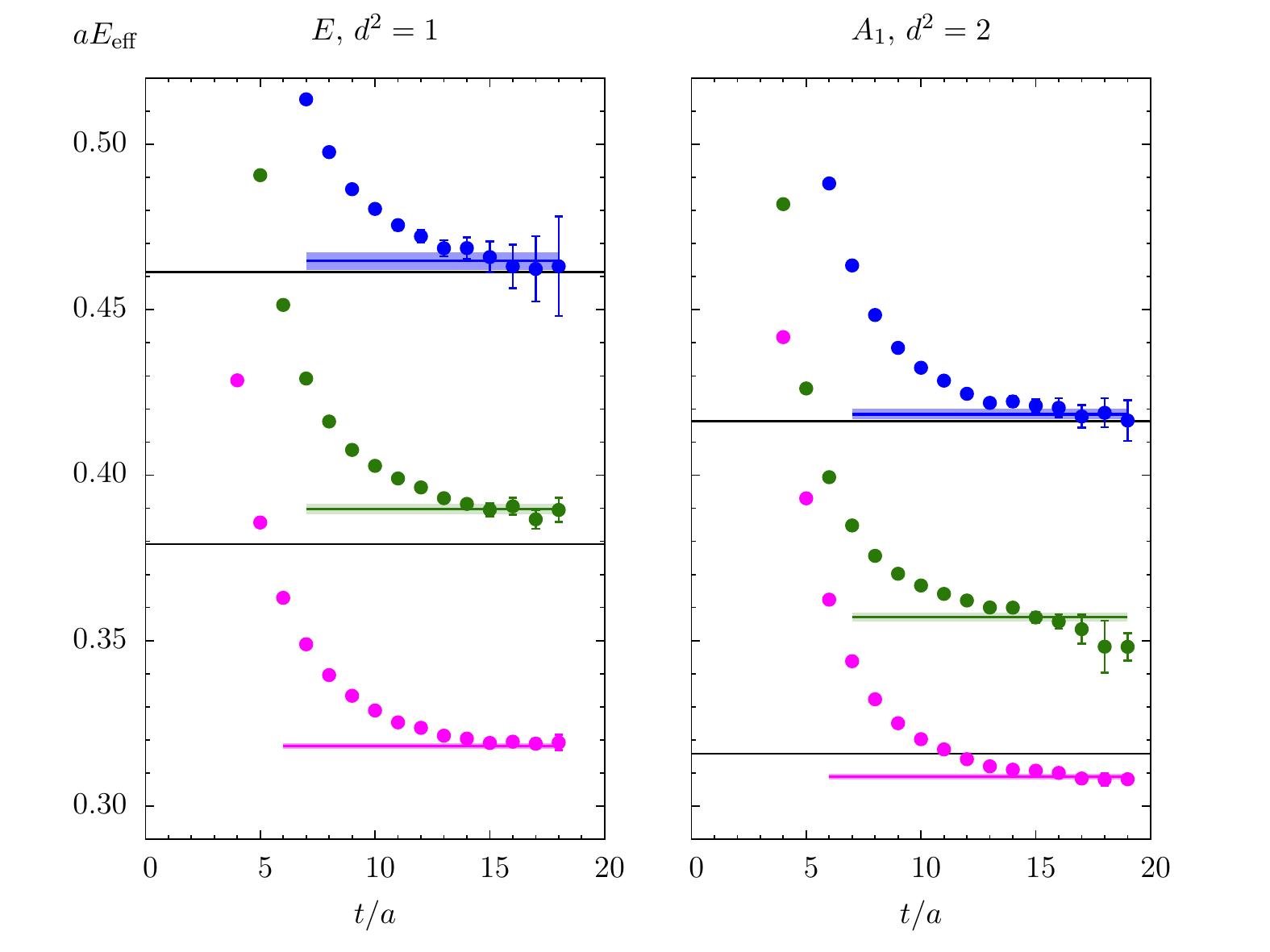}
\caption{Spectrum from the GEVP on the F6 lattice, using the window
  method, for two irreps: $\mathbf{d}^2=1, E$ on the left (a typical
  example for the energy levels we extract) and $\mathbf{d}^2=2, A_1$
  on the right (an example where the plateau does not look as good,
  particularly the intermediate level). The different levels in the
  respective irrep are plotted using different colours and the
  accordingly coloured bands are the fit results of the corresponding
  eigenvalues to fit function allowing for the ground state and an
  excited state. The width of those bands indicates the statistical
  error of the fit and the length shows the chosen fit range. The
  horizontal lines are the free two-pion levels in the respective
  moving frame.}
  \label{figure:spectrum-F6-window}
\end{figure}

The pion masses on the three ensembles have been extracted using a $\cosh$-fit ansatz. The fit ranges and results are shown in Table \ref{table:pion-fits}.
\begin{table}[tb]
\centering
\begin{tabular}{|c|c|c|c|}
\hline 
 & $t_\mathrm{min}$ & $t_\mathrm{max}$ &  $a m_\pi$\\ 
\hline 
E5 & $15$ & $28$ & $0.14511(33)$\\ 
\hline 
F6 & $16$ & $35$ & $0.10366(29)$\\ 
\hline 
F7 & $19$ & $40$ & $0.08893(30)$\\
\hline
\end{tabular}
\caption{Fit ranges to the single-cosh fit and corresponding pion masses including jackknife error on the three ensembles.}
\label{table:pion-fits}
\end{table}
We also solved the GEVP in the window method for the 8 irreps listed in
Table~\ref{table:irreps-rho-pipi}. The extracted energy levels of two
  selected irreps on the F6 lattice, together with the effective energies
  $E^{(k)}_{\mathrm{eff}}(t)=-t_w^{-1} \ln( \lambda^{(k)}(t))$ are shown in Fig. \ref{figure:spectrum-F6-window}.

The energy levels were obtained by fitting the eigenvalues extracted
from the GEVP to a function allowing for the ground state and one excited state. Results
for all irreps and ensembles are listed in
Table~\ref{table:energy-spectrum-window}, together with the values of
$\chi^2$/d.o.f. for each fit.

\begin{table}[htbp]
\centering

\begin{tabular}{cc|cc|cc|cc}
\hline
\hline
$d^2$ & irrep & E5 & $\chi^2$/d.o.f. & F6 & $\chi^2$/d.o.f. & F7 & $\chi^2$/d.o.f.   \\
\hline
& & 0.3213(11)& 0.77& 0.2883(9)& 0.63& 0.2727(11)& 0.45\\
0 &$T_1$ &0.4905(21)& 0.73& 0.3443(15)& 0.82& 0.3306(17)& 1.67\\
& & & & 0.4333(32)& 0.42& 0.4228(34)& 0.75\\
\hline
& & 0.3022(8)& 1.05& 0.2329(10)& 0.84& 0.2049(8)& 1.40\\
1 &$A_1$ &0.3573(12)& 1.12& 0.2996(15)& 0.96& 0.2875(18)& 0.66\\
& & & & 0.3618(18)& 1.24& 0.3491(21)& 1.04\\
\hline
& & 0.3215(14)& 1.67& 0.2900(10)& 0.83& 0.2755(11)& 1.14\\
1 &$E$ &0.5238(41)& 0.77& 0.3671(18)& 1.02& 0.3559(18)& 0.49\\
& & & & 0.4460(28)& 0.35& 0.4356(40)& 0.56\\
\hline
& & 0.3068(11)& 0.85& 0.2472(10)& 1.18& 0.2224(11)& 0.95\\
2 &$A_1$ &0.3783(20)& 1.37& 0.3054(17)& 1.23& 0.2945(21)& 1.04\\
& & & & 0.3753(18)& 0.61& 0.3646(21)& 1.77\\
\hline
& & 0.3155(25)& 0.60& 0.2658(10)& 0.86& 0.2467(11)& 1.82\\
2 &$B_1$ &0.4128(20)& 1.11& 0.3106(18)& 1.25& 0.2948(29)& 1.67\\
& & & & 0.3841(20)& 1.48& 0.3700(27)& 0.43\\
\hline
& & 0.3240(23)& 1.10& 0.2913(13)& 0.84& 0.2783(17)& 2.22\\
2 &$B_2$ &0.5454(51)& 1.32& 0.3755(18)& 1.29& 0.3653(20)& 0.71\\
& & & & 0.3943(24)& 1.06& 0.3797(40)& 0.78\\
\hline
& & 0.3096(18)& 0.59& 0.2584(13)& 0.37& 0.2364(17)& 1.35\\
3 &$A_1$ &0.3937(47)& 0.62& 0.2989(14)& 0.44& 0.2831(20)& 0.82\\
& & & & 0.3161(21)& 0.97& 0.3079(26)& 0.82\\
\hline
3 &$E$ &0.3199(37)& 1.83& 0.2786(12)& 1.41& 0.2617(15)& 0.73\\
& & 0.4538(37)& 0.77& 0.3295(18)& 1.12& 0.3132(34)& 1.74\\
\hline
\hline
\end{tabular} 
\caption{Extracted energy levels $a E_k$ (states are ordered from ground state to the higher excited states from top to bottom) in the window method with $t_w$=3 in each irrep for the three lattices used in this work. One level fewer per irrep is extracted on E5, due to the levels being above the $4m_\pi$ threshold and the interpolator basis being smaller by 1.}
\label{table:energy-spectrum-window}
\end{table}

\subsection{L\"uscher formalism}

L\"uscher's finite volume method \cite{luscher1,luscher2} is used to
map the energy levels of the finite-volume lattice box to the continuum phase
shift.\footnote{For a review of recent physics results from (extensions of)
  the L\"uscher method see Ref. \cite{Briceno:2017max}.} For the
$\rho$, we are interested in the $l=1$ partial wave. In principle, higher
partial waves also contribute to the spectrum. The effect of the $l=3$ and $l=5$ partial
waves has been studied in \cite{irrep-zetas,Andersen:2018mau}. With this restriction, the quantisation condition reads
\begin{align}
\delta_1(k) = \phi^{\mathbf{d}}_\Lambda(q) + n \pi
\, .
\label{luescher}
\end{align}
In this equation, $k = ( 2 \pi / L ) q$ are the scattering momenta,
$\delta_1(k)$ is the $l=1$ infinite volume phase shift, and $\phi^{\mathbf{d}}_\Lambda(q)$ is a kinematical function related to modified
zeta functions, which can be computed to arbitrary precision. The centre-of-mass energy is given by $E_{\textrm{cm}} = 2\sqrt{m_\pi^2 + k^2}$. With the spectrum data from the GEVP we can use this relation to map out the
infinite volume phase shift in the energy region $2 m_\pi < E < 4 m_\pi$. 

\begin{figure}[htbp] % no figure before 1st section
  \centering
\includegraphics[width=.99\linewidth,clip]{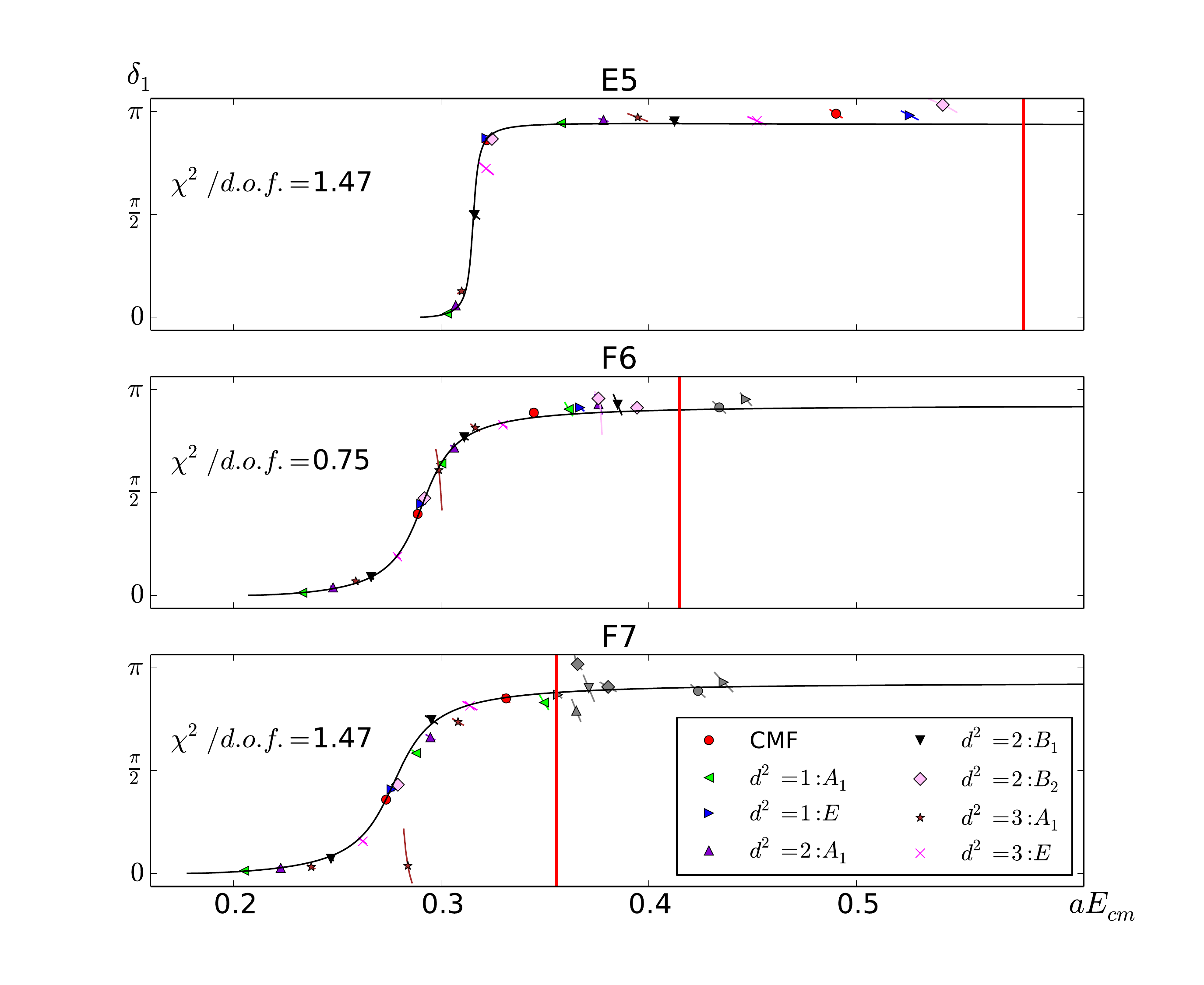}
  \caption{Phase shifts on all three ensembles using the window method. The horizontal axis shows the CMF energy of each level and data points of the same colour and symbol belong to the same frame and irrep. Error bars follow the curves allowed by the L\"uscher zeta functions. The red vertical line indicates the $4 m_\pi$ threshold in each system; data points above are excluded from the fit and thus shown in grey. The black line is the result of the Breit-Wigner fit to our data by minimising the $\chi^2$ functional defined in \eqref{eqn:chi2-fit-delta}. The $\chi^2$/d.o.f. for each fit is shown in the plots.}
  \label{figure:delta-window}% Give a unique label
\end{figure}

The results from this procedure are shown in Fig.~\ref{figure:delta-window} for all three ensembles used.
The curve in this plot is a fit to a Breit-Wigner parameterisation,
\begin{align}
k^3\cot\delta_1^\mathrm{BW}(k;g_{\rho\pi\pi},m_\rho) &= \frac{6 \pi}{g_{\rho\pi\pi}^2} (m_\rho^2 - E_{\textrm{cm}}^2) E_{\textrm{cm}} 
\, ,
\label{eqn:breit-wigner}
\end{align}
which is motivated in the resonance region by the effective-range
formula. Given that the data points and their error estimates are confined to
the curves dictated by the L\"uscher zeta function, as is visible in Fig.~\ref{figure:delta-window}, we fit the data according to their error
behaviour along the curves dictated by the zeta-functions. The L\"uscher
condition is reformulated to 
\begin{align}
\cot \delta_1(k) = \cot (\phi(q))\,,
\end{align}
and the difference to Equation \eqref{eqn:breit-wigner},
\begin{align}
f(q;g_{\rho\pi\pi},m_\rho) = \cot ( \phi(q)) - \cot\delta_1^\mathrm{BW} \bigg(\frac{2 \pi}{L} q;g_{\rho\pi\pi},m_\rho \bigg)
\, ,
\end{align}
is calculated. Given any pair of resonance parameters $(g_{\rho\pi\pi},m_\rho)$ we can solve $f(q;g_{\rho\pi\pi},m_\rho) = 0$, and this way obtain $q_i^\textrm{fit}(g_{\rho\pi\pi},m_\rho)$ and the energy levels $E_{\textrm{cm},i}^\textrm{fit}(g_{\rho\pi\pi},m_\rho)$. To this end define the $\chi^2$-function
\begin{align}
\chi^2(g_{\rho\pi\pi},m_\rho) &= \sum_{i,j} (E_{\textrm{cm},i}^\textrm{fit}(g_{\rho\pi\pi},m_\rho) - E_{\textrm{lat},i}) C^{-1}_{i,j} 
(E_{\textrm{cm},j}^\textrm{fit}(g_{\rho\pi\pi},m_\rho) - E_{\textrm{lat},j})
\, ,
\label{eqn:chi2-fit-delta}
\end{align}
with the covariance matrix
\begin{align}
C_{i,j}  &=  \sum_{k=0}^{n_\textrm{jk}} (E_{\textrm{lat},i,k} - \bar{E}_{\textrm{lat},i}) (E_{\textrm{lat},j,k} - \bar{E}_{\textrm{lat},j})
\, ,
\end{align}
calculated from the $n_\textrm{jk}$ jackknife samples of the lattice energies
and their central values $\bar{E}_{\textrm{lat},i}$. By
minimising this $\chi^2$-function on each jackknife sample, we can obtain fit
values for the resonance parameters.  One advantage of this approach is
that we can use any parameterisation suitable for the situation. We can compare our
form factor results to the Gounaris-Sakurai parameterisation
\cite{gounaris-sakurai} of the phase shift, which is characterised by the resonance parameters $m_\rho$ and $\Gamma_\rho = \frac{k_\rho^3}{m_\rho^2} \frac{g_{\rho\pi\pi}^2}{6 \pi}$:
\begin{gather}
\frac{k^3}{E_\mathrm{cm}} \cot[\delta_{1}^\mathrm{GS}(k)] = k^2 h(E_\mathrm{cm}) - k^2_\rho h(m_\rho) + (k^2 - k_\rho^2) b
\, ,
\label{GS-delta} \\
b= -\frac{2}{m_\rho} \bigg[ \frac{2 k_\rho^3}{m_\rho \Gamma_\rho} + \frac12 m_\rho h(m_\rho) +k_\rho^2 h'(m_\rho)\bigg]
\, , \\
h(\omega) = \frac{2k_\omega}{\pi \omega} \ln\frac{\omega + 2k_\omega}{2 m_\pi}
\, , \\
k_\omega = \sqrt{\frac{\omega^2}{4}-m_\pi^2}\, ,\quad k_\rho = k_{m_\rho}
\, ,
\label{GS-final}
\end{gather} 
and show the results in Table \ref{table:phase-shift-window}. The
  two fits produce consistent results, although the Gounaris-Sakurai
  parameterisation yields slightly higher values of $\chi^2$.
\begin{table}[tbp]
\centering
\begin{tabular}{c|cc|cc|cc}
\hline
\hline
  & \multicolumn{2}{c}{E5}  & \multicolumn{2}{c}{F6} & \multicolumn{2}{c}{F7}   \\
\hline
  & BW & GS & BW & GS & BW & GS   \\
\hline
$m_\rho$ & 0.3156(8) & 0.3157(10) & 0.2933(8) & 0.2934(9) & 0.2800(10) & 0.2800(10)\\
$g_{\rho\pi\pi}$ & 5.70(9) & 5.66(9) & 6.08(13) & 6.03(13) & 5.91(17) & 5.88(16)\\
$\chi^2$/d.o.f. & 1.47 & 1.64 & 0.75 & 0.84 & 1.47 & 1.52\\
\hline
\end{tabular} 

\caption{Resonance parameters extracted from the fit to the energy
  levels using the L\"uscher formalism. All levels are extracted using
  the window method with $t_w$=3. Compared are the fit results to the
  Breit-Wigner and Gounaris-Sakurai parameterisations.}
\label{table:phase-shift-window}
\end{table}

\begin{figure}[tbp] % no figure before 1st section
  \centering
\includegraphics[width=1.0\linewidth,clip]{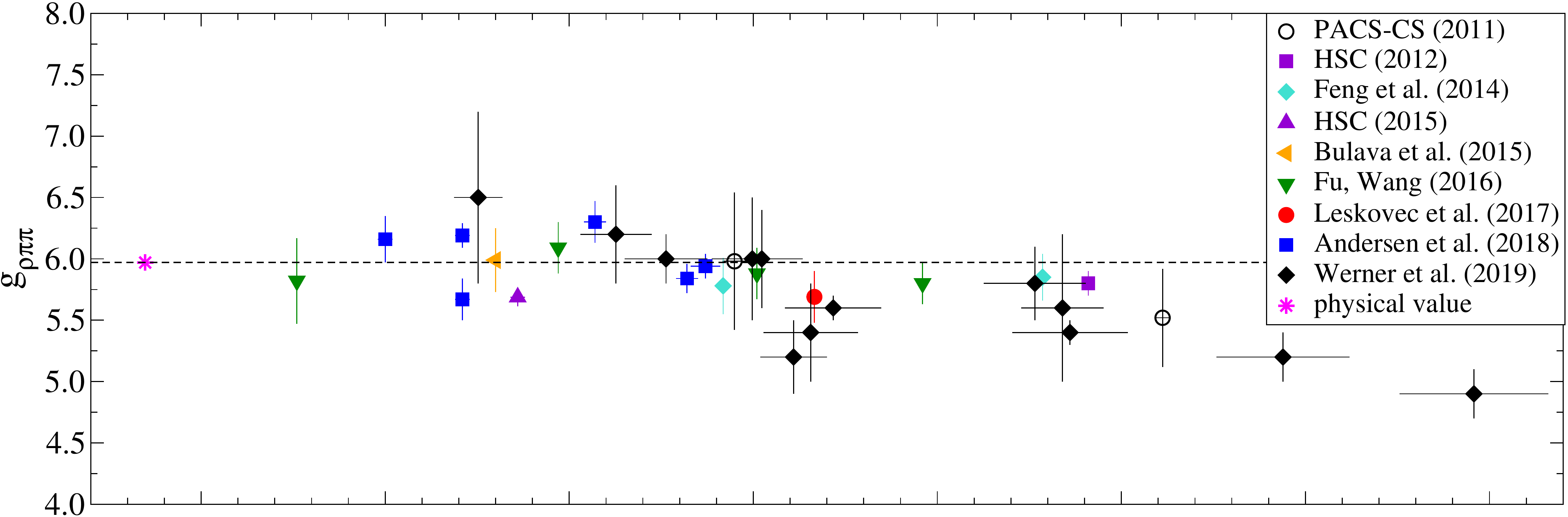}\\
\includegraphics[width=1.0\linewidth,clip]{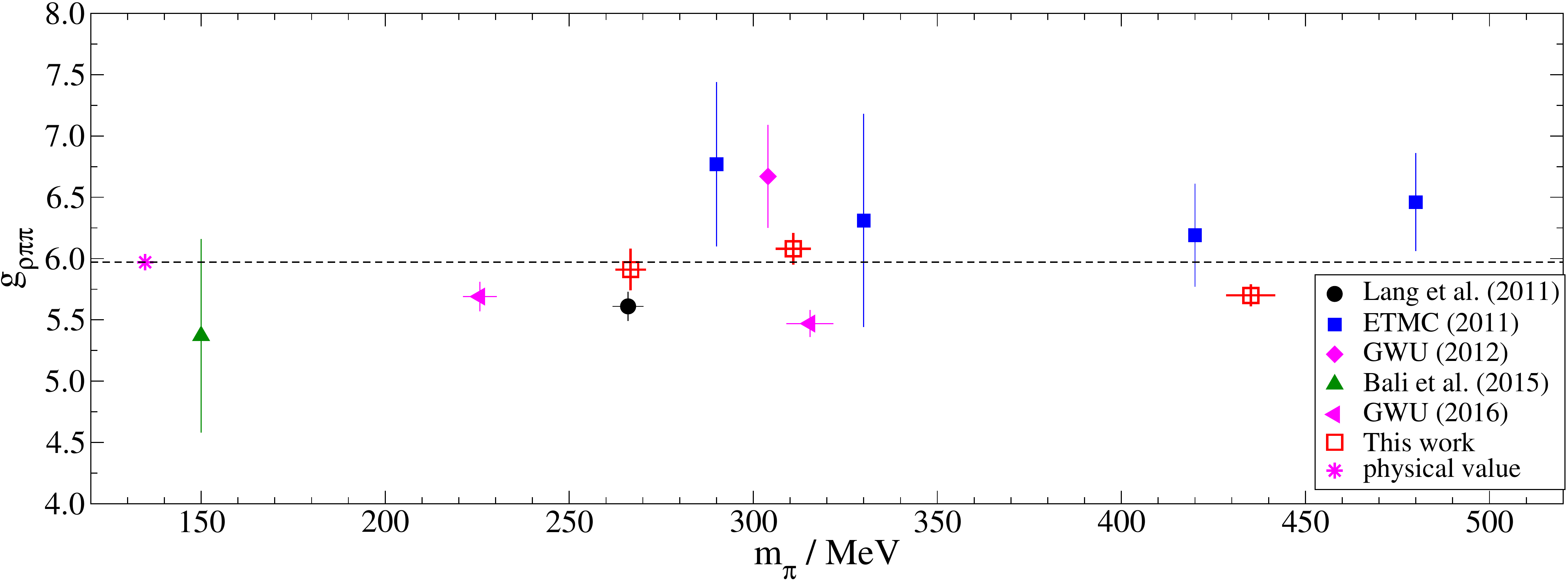}
  \caption{Overview of lattice results for the coupling $g_{\rho\pi\pi}$ as a
    function of the pion mass in the calculation. The upper pane shows the
    results from simulations with dynamical light and strange quarks, while
    the lower pane shows results with dynamical light quarks only. Where
    available, the scale-setting uncertainty provided by the authors has been
    added in quadrature to obtain the errors on the horizontal axis. The value extracted from the physical $\rho$-meson width is indicated by the magenta star and the black dashed line. The
    results from this work are the red open squares in the lower pane.}
  \label{overview}% Give a unique label
\end{figure}

Figure \ref{overview} shows the world data for the coupling
$g_{\rho\pi\pi}$ from various 2 and 2+1 flavor simulations. There
  is no significant dependence on the pion mass, and the lattice
  results are generally close to the physical value.

\begin{table}[htbp]
\centering
\begin{tabular}{cc|cc|cc|cc}
\hline
\hline
$d^2$ & irrep & \multicolumn{2}{c}{E5}  & \multicolumn{2}{c}{F6}  & \multicolumn{2}{c}{F7}  \\ 
\hline
$Z_V$ &  & \multicolumn{2}{c}{$0.74418(33)$}  & \multicolumn{2}{c}{$0.74143(14)$}  & \multicolumn{2}{c}{$0.74011(23)$}  \\ 
\hline
 &  & $|A_l|$ & $|A_c| $  & $|A_l|$ & $|A_c| $  & $|A_l|$ & $|A_c|$    \\ 
\hline
& & 2.41(34)& 2.12(31)& 1.94(22)& 1.74(19)& 1.79(24)& 1.63(22)\\
0 &$T_1$ &0.75(20)& 0.57(16)& 1.06(18)& 0.90(16)& 1.05(17)& 0.90(15)\\
& & & & 0.71(22)& 0.53(18)& 0.80(20)& 0.65(18)\\
\hline
& & 2.02(25)& 1.81(22)& 0.55(7)& 0.51(6)& 0.48(6)& 0.46(5)\\
1 &$A_1$ &1.85(25)& 1.59(22)& 2.18(30)& 1.95(26)& 2.08(30)& 1.87(26)\\
& & & & 0.79(14)& 0.66(12)& 0.87(15)& 0.74(13)\\
\hline
& & 2.24(33)& 1.98(30)& 1.94(21)& 1.74(19)& 1.79(23)& 1.63(21)\\
1 &$E$ &1.18(35)& 0.97(30)& 0.89(16)& 0.74(14)& 0.98(16)& 0.82(14)\\
& & & & 0.57(17)& 0.42(14)& 0.52(14)& 0.40(11)\\
\hline
& & 2.44(32)& 2.17(28)& 0.83(10)& 0.77(9)& 0.72(10)& 0.68(9)\\
2 &$A_1$ &1.57(26)& 1.33(23)& 2.25(32)& 2.00(28)& 2.22(33)& 1.98(29)\\
& & & & 0.65(12)& 0.54(10)& 0.63(11)& 0.53(10)\\
\hline
& & 2.05(39)& 1.81(35)& 1.02(12)& 0.93(11)& 0.82(11)& 0.76(11)\\
2 &$B_1$ &0.98(17)& 0.81(15)& 1.77(25)& 1.56(22)& 1.76(30)& 1.57(26)\\
& & & & 0.43(10)& 0.33(9)& 0.50(10)& 0.41(9)\\
\hline
& & 2.18(41)& 1.92(37)& 1.91(24)& 1.71(21)& 1.77(27)& 1.60(24)\\
2 &$B_2$ &0.61(19)& 0.50(16)& 0.38(8)& 0.31(7)& 0.16(4)& 0.14(4)\\
& & & & 0.92(18)& 0.74(15)& 0.87(20)& 0.69(16)\\
\hline
& & 2.75(44)& 2.44(40)& 1.17(15)& 1.08(13)& 1.01(17)& 0.94(16)\\
3 &$A_1$ &1.41(31)& 1.17(26)& 1.27(18)& 1.12(15)& 0.98(14)& 0.87(13)\\
& & & & 1.87(29)& 1.63(26)& 2.05(33)& 1.81(29)\\
\hline
3 &$E$ &1.42(33)& 1.25(30)& 1.40(18)& 1.26(15)& 1.25(18)& 1.14(17)\\
& & 0.78(17)& 0.64(14)& 1.40(22)& 1.20(19)& 1.40(25)& 1.21(22)\\
\hline
\hline
\end{tabular}
\caption{Matrix elements $|A_{l/c}|$ extracted from the window method in units
  of $10^{-2}$. The values for $Z_V$ are taken from \cite{mainz:g-2}. The
  difference in $|A_l|$ and $|A_c|$ is likely due to cut-off
  effects, which are studied in \cite{mainz:g-2, gerardin:2019rua}.}
\label{table:a-psi-window}
\end{table}

\section{The timelike pion form factor}
\label{tff}
To determine the timelike pion form factor we first need to calculate the
matrix elements $|A_{l/c}|_n = |\langle 0 | J_{l/c} | n \rangle|$.
The subscripts $l/c$ refer to the local and the conserved currents, respectively. Our results for the matrix elements
$|A_{l/c}|$ are listed in Table \ref{table:a-psi-window}. There are sizable
differences between $|A_l|$ and $|A_c|$, likely due to cut-off effects, which
are studied in \cite{mainz:g-2, gerardin:2019rua}. This is a clear indication that an improved version of the currents (defined e.g. in \cite{impr-curr-1,impr-curr-2}) would be preferable. These differences are supposed to vanish in the continuum limit, but we cannot check this since we are only considering a single lattice spacing.

\label{timelike}
We now have all the input to compute the timelike pion form factor \cite{Meyer:timelike,timelike-ff-mf},
\begin{align}
|(F_\pi)^{\mathbf{d}}_\Lambda(s)|^2 = G^{\mathbf{d}}_\Lambda(\gamma) \Big( q (\phi^{\mathbf{d}}_\Lambda)'  (q) + k \frac{\partial \delta_1(k)}{\partial k} \Big) \frac{3 \pi s}{k^5} |A|^2
\, ,
\label{eqn:pion-form-factor}
\end{align}
where $s=E_\mathrm{cm}^2$ and
\begin{align}
G^{\mathbf{d}}_\Lambda(\gamma)=
\begin{cases} 
\frac 1\gamma \phantom{aa} \mathrm{if} \phantom {aa} \Lambda = A_1 \\
 \gamma \phantom{aa} \mathrm{otherwise}
 \end{cases}
\, ,
\end{align}
with the Lorentz-boost $\gamma = \frac{E}{E_\mathrm{cm}}$.

This equation includes derivatives of the infinite volume phase shift
$\delta_1(k)$ obtained in the previous section, and of the modified L\"uscher
zeta functions $\phi^{\mathbf{d}}_\Lambda$, which were also used in the
phase-shift analysis, and which can be obtained to any desired mathematical
precision. We want to compare our form factor results to another study
\cite{mainz:g-2}, which was performed on the same ensembles and which used
correlators with one local and one conserved vector current. To conform with that study, we define the local-conserved version of $|A|^2$,
\begin{align}
|A_{lc}|^2 \equiv |A_l| |A_c|
\, .
\label{eqn:apsi-lc}
\end{align}
For another literature comparison \cite{mainz:sq-rad}, we use the local-local version.

\begin{figure}[htbp] % no figure before 1st section
  \centering
\includegraphics[width=0.8\linewidth,clip]{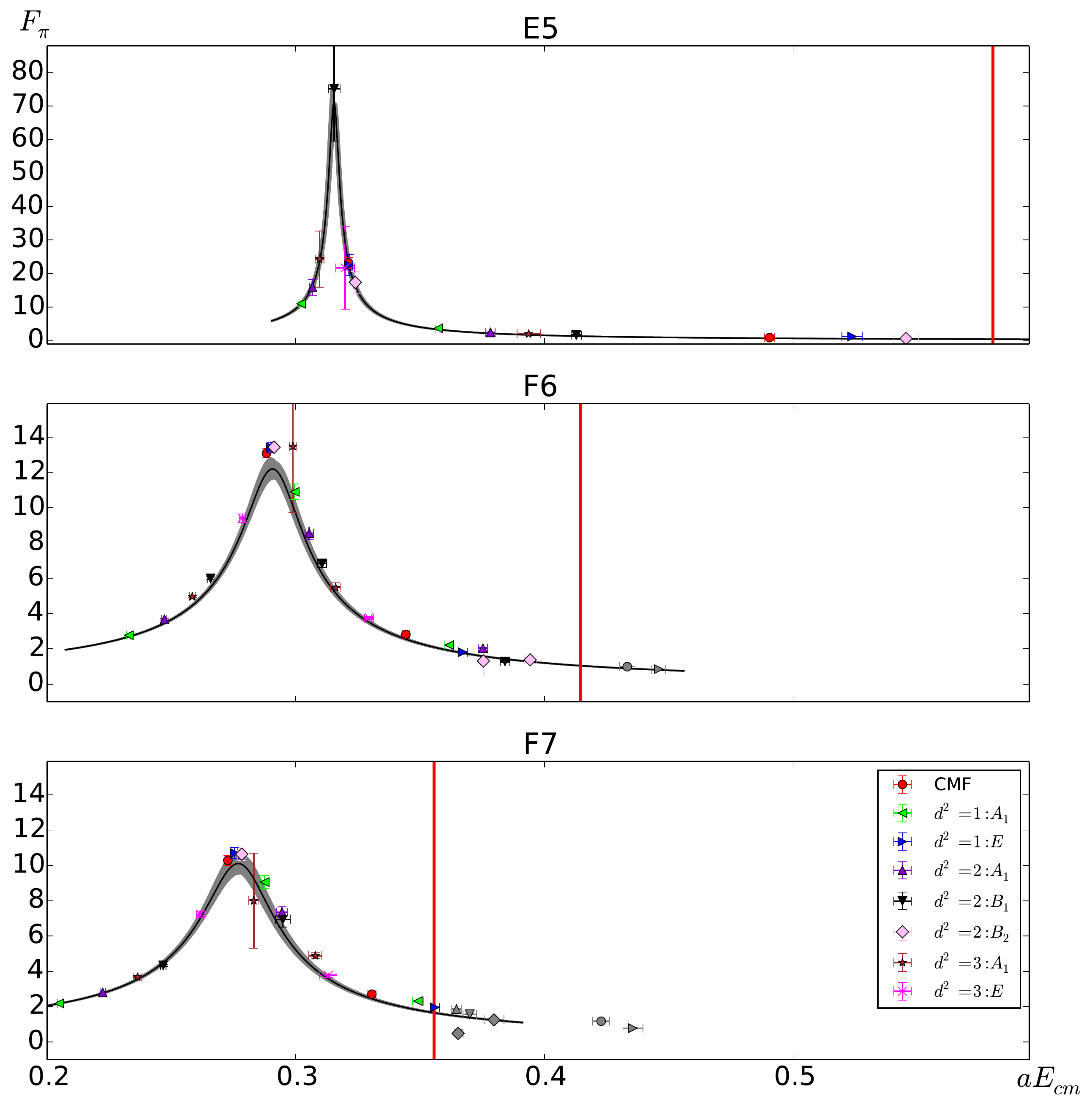}
  \caption{The timelike pion form factor on the E5, F6, F7 ensembles (top to bottom). Data points with the same symbol and colour belong to the same frame and irreps. The error bars associated with each data point come from a jackknife estimate. The grey curve is the GS representation of $F_\pi$, which only takes the fit parameters of the phase-shift fit $m_\rho, g_{\rho\pi\pi}$ into account --- it is not a fit to the data pictured in these plots. The vertical red bars indicate the $4 m_\pi$ threshold for each lattice.}
  \label{figure:Fpi-GS-window}% Give a unique label
\end{figure}

Equation \eqref{eqn:pion-form-factor} allows us to directly determine
$F_\pi(s)$ from lattice data for discrete values of $s$, using a
parameterisation of the phase shift as well as the current matrix elements. To get a continuous description of $F_\pi(s)$, we can use the Gounaris-Sakurai parameterisation
\cite{gounaris-sakurai}, given by the resonance parameters $m_\rho, \Gamma_\rho$:
\begin{align}
F_\pi^{\mathrm{GS}}(E_\mathrm{cm}) &= \frac{f_0}{\frac{k^3}{E_\mathrm{cm}}(\cot[\delta_{1}^{\mathrm{GS}}(k)]-i)} \, ,\\
f_0 &= -\frac{m_\pi^2}{\pi} - k_\rho^2 h(m_\rho)-b \frac{m_\rho^2}{4}
\, ,
\end{align}
with the definitions from Eqs. (\ref{GS-delta} -- \ref{GS-final}).
The comparison of our lattice-calculated values for $F_\pi$ and the
Gounaris-Sakurai curves is shown in Figure \ref{figure:Fpi-GS-window}.  We
want to stress these these curves are not fits to the form factor data.

\begin{figure}[htbp] % no figure before 1st section
  \centering
\includegraphics[width=0.99\linewidth,clip]{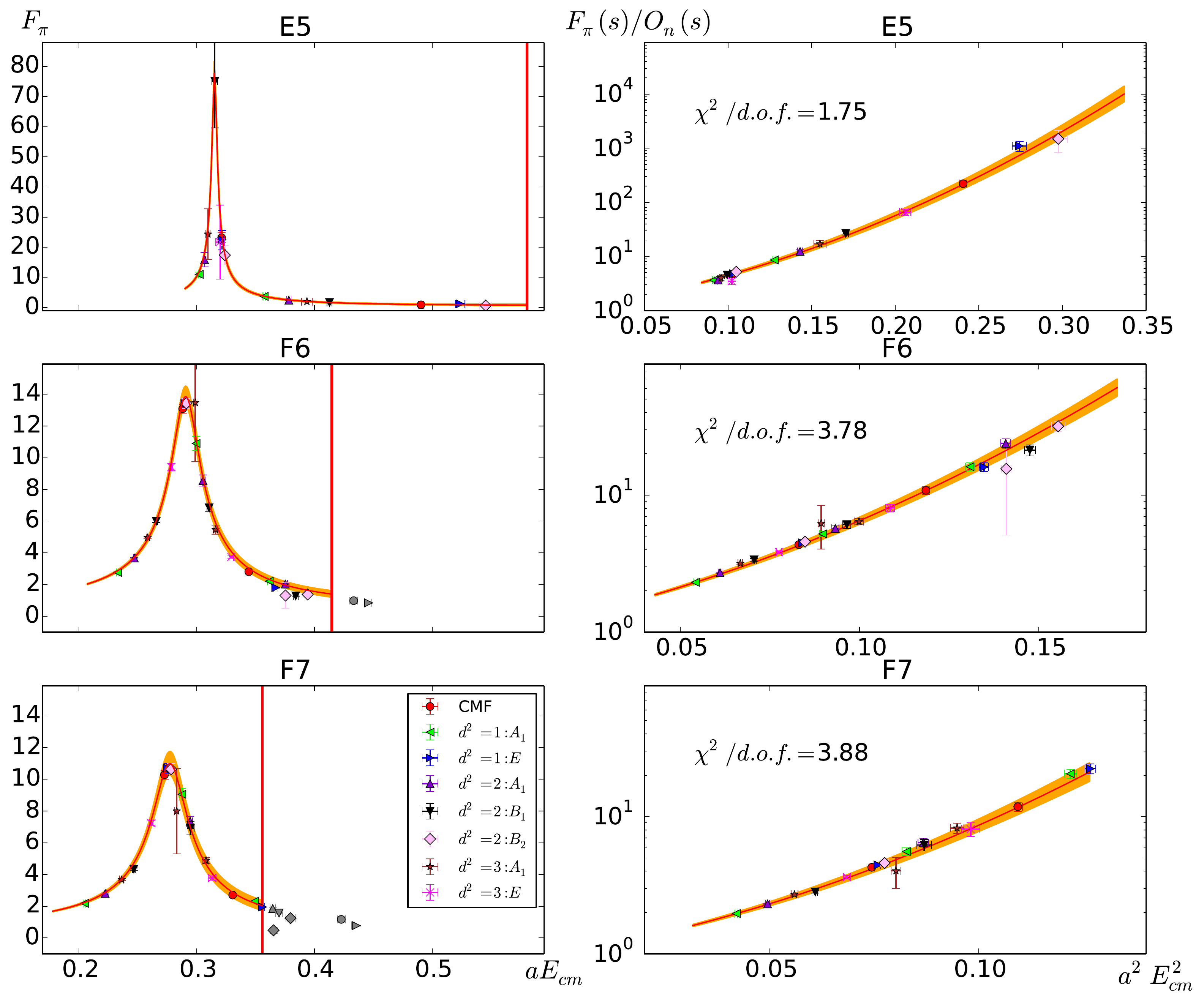}
  \caption{\textbf{Left panel: }The timelike pion form factor on the E5, F6, F7
  lattice (top to bottom), window method. Data points with the same symbol and
  colour belong to the same frame and irreps. The orange curve is the fit to
  $F_\pi$, parametrised via the $3$-subtracted version of Equation
  \eqref{omnes}. The vertical red bars indicate the $4 m_\pi$ threshold in
  each lattice and data points above this threshold have not been included in
  the fit and are shown in grey for this reason. \textbf{Right panel: } The
  data which we are actually fitting to. The vertical axis shows $F_\pi$ divided by
  the Omn\`es integral, i.e. the analytically calculable part of Equation \eqref{omnes}, and the fit function is
  $f(s)=\exp(Ps)$, where P is a 1st-order polynomial. The vertical axis is displayed on
  a $\log$ scale and the orange curve is the fit function with the jackknife
  error. Shown are also the $\chi^2 / \mathrm{d.o.f.}$ values of the respective
  fits, which are quite high.}
  \label{figure:Fpi-window-3}% Give a unique label
\end{figure}

The Gounaris-Sakurai curve seems to describe our data reasonably well, but it
would be desirable to have a fit to our form factor data extracted from
lattice QCD. One way to realise such a fit is an $n$-subtracted Omn\`{e}s representation \cite{form-factor-streuphase,form-factor-streuphase2}
\begin{align}
F_\pi^{(n)}(s) = \exp \Bigg( P_{n-1}(s) s + \frac{s^n}{\pi} \int_{4 m_\pi^2}^\infty ds' \frac{\delta_{1}(s')}{(s')^{n}(s'-s-i\epsilon)} \Bigg)
\, ,
\label{omnes}
\end{align}
where $P_{n-1}(s)$ is a polynomial function of degree $n-1$. We parametrise the phase shift $\delta_{1}(s')$ in this equation using the Breit-Wigner form, Equation \eqref{eqn:breit-wigner}, and our extracted resonance parameters. For the $2$-subtracted version, the polynomial is a constant,
\begin{align}
P_1(s) = \frac{\langle r_\pi^2 \rangle}{6}
\, ,
\end{align}
with the square radius $\langle r_\pi^2 \rangle$ of the pion. The polynomial for the $3$-subtracted version reads
\begin{align}
P_2(s) = \frac{\langle r_\pi^2 \rangle}{6} + \frac{1}{2} \Big(2 c_V^\pi - \Big( \frac{\langle r_\pi^2 \rangle}{6} \Big)^2 \Big) s
\, ,
\label{omnes-fit}
\end{align}
with the curvature $c_V^\pi$ of the pion form factor. The integrand of
\begin{align}
O_n(s) = \exp \Bigg( \frac{s^n}{\pi} \int_{4 m_\pi^2}^\infty ds' \frac{\delta_{1}(s')}{(s')^{n}(s'-s-i\epsilon)} \Bigg)
\end{align}
has a pole at $s'=s$ and in order to solve the integral numerically we need to do a subtraction,
\begin{align}
\int_{4 m_\pi^2}^\infty ds' \frac{\delta_{1}(s')}{(s')^n(s'-s-i\epsilon)} = \int_{4 m_\pi^2}^\infty ds' \frac{\delta_{1}(s') - \delta_{1}(s)}{(s')^n(s'-s)} + \delta_{1}(s) \int_{4 m_\pi^2}^\infty ds' \frac{1}{(s')^n(s'-s-i\epsilon)}
\, .
\label{modified}
\end{align}
The integral $O(s)$ can now be computed analytically. We divide the lattice data $F_\pi(s)$ by the function $O_n(s)$, and fit the result using the function
$f_\mathrm{fit}(s) = \exp(P_{n-1}(s)s)$. The results of the fit to the $3$-subtracted version are
shown in Figure \ref{figure:Fpi-window-3}. In the $2$-subtracted version, our
data were not very well described by the fit function. The fit describes the
$F_\pi$ data much better than the GS representation of the form factor, but for all ensembles the fits have somewhat large values for
$\chi^2 / \mathrm{d.o.f.}$ We investigated the cause of this and excluded
  autocorrelation in the chain or single outlying data points as sources for
  this observation. There are however indications that our data set might be
  too small for reliable estimates of such a large covariance matrix. 
  
Results for the square radius $\langle r_\pi^2 \rangle$ from this fit are shown in Table \ref{table:r2-window-lc}. The results for the $2$- and
$3$-subtracted version differ on the level of $2 \sigma$, which is another indication that the $2$-subtracted version is not enough to describe the data
accurately. The square radius was previously determined in \cite{mainz:sq-rad} by
fitting the spacelike pion form factor, computed on the same ensembles we are using in our study. This is a completely different approach and provides a very good cross-check of our fit procedure. Because the authors of
\cite{mainz:sq-rad} employ a local current (as opposed to the local-conserved
setup used up to this point), we repeated the analysis using
$|A_l|$ in Equation \eqref{eqn:pion-form-factor}. The results
for the square radius from this analysis are shown and compared to the result
from \cite{mainz:sq-rad} in Table \ref{table:r2-window-ll}. While both
  results agree very well for ensembles E5 and F6, we obtain a somewhat
  smaller square radius on ensemble F7. This observation is discussed further
  in Section \ref{hvp}. The comparison of
this table with Table \ref{table:r2-window-lc} shows again that discretisation
effects in our currents are sizable.

\begin{table}[tbp]
\centering
\begin{tabular}{cc|c|c|c}
\hline
\hline
 & $n$ &E5  & F6  & F7  \\ 
\hline
$\langle r_\pi^2 \rangle /r_0^2$  & 2 &1.18(2) & 1.34(1) & 1.46(2)\\
\hline
$\langle r_\pi^2 \rangle/r_0^2$  &3 &1.11(3) & 1.31(3) & 1.37(4)\\
$c_V/r_0^4 $ & 3 &3.59(7) & 4.98(7) & 6.05(15)\\
\hline
\end{tabular}
\caption{Square radius and curvature (in units of $10^{-2}$) of the pion
  obtained from the fit to the $n$-subtracted Omn\`es representation of the
  form factor, using a local-conserved current setup. The Sommer scale $r_0$
  is taken from \cite{Fritzsch:2012wq}. We want to stress that the curvature
  can indeed be calculated from the fit parameter we use, but that the result
  and particularly the error estimate presented here might not be the physical
  value.}
\label{table:r2-window-lc}
\end{table}
\begin{table}[tbp]
\centering
\begin{tabular}{cc|c|c|c}
\hline
\hline
 & $n$ &E5  & F6  & F7  \\ 
\hline
$\langle r_\pi^2 \rangle /r_0^2$  & 2 &1.25(2) & 1.41(1) & 1.53(2)\\
\hline
$\langle r_\pi^2 \rangle/r_0^2$  &3 &1.18(3) & 1.37(3) & 1.43(4)\\
$c_V/r_0^4 $ & 3 &3.81(7) & 5.26(8) & 6.33(15)\\
\hline
$\langle r_\pi^2 \rangle /r_0^2$ \cite{mainz:sq-rad} & & 1.18(5) & 1.37(6) & 1.61(10) \\
\hline
\end{tabular}
\caption{Same as Table \ref{table:r2-window-lc} but using a local-local vector
  current. The last line shows the values from \cite{mainz:sq-rad}, where
  $\langle r_\pi^2 \rangle$ has been computed from a fit to the spacelike form factor. The difference of our results to the corresponding values in Table \ref{table:r2-window-lc} comes from discretisation effects, which are also visible in the matrix elements themselves, shown in Table \ref{table:a-psi-window}.}
\label{table:r2-window-ll}
\end{table}

\begin{figure}[htbp] % no figure before 1st section
  \centering
\includegraphics[width=0.49\linewidth,clip]{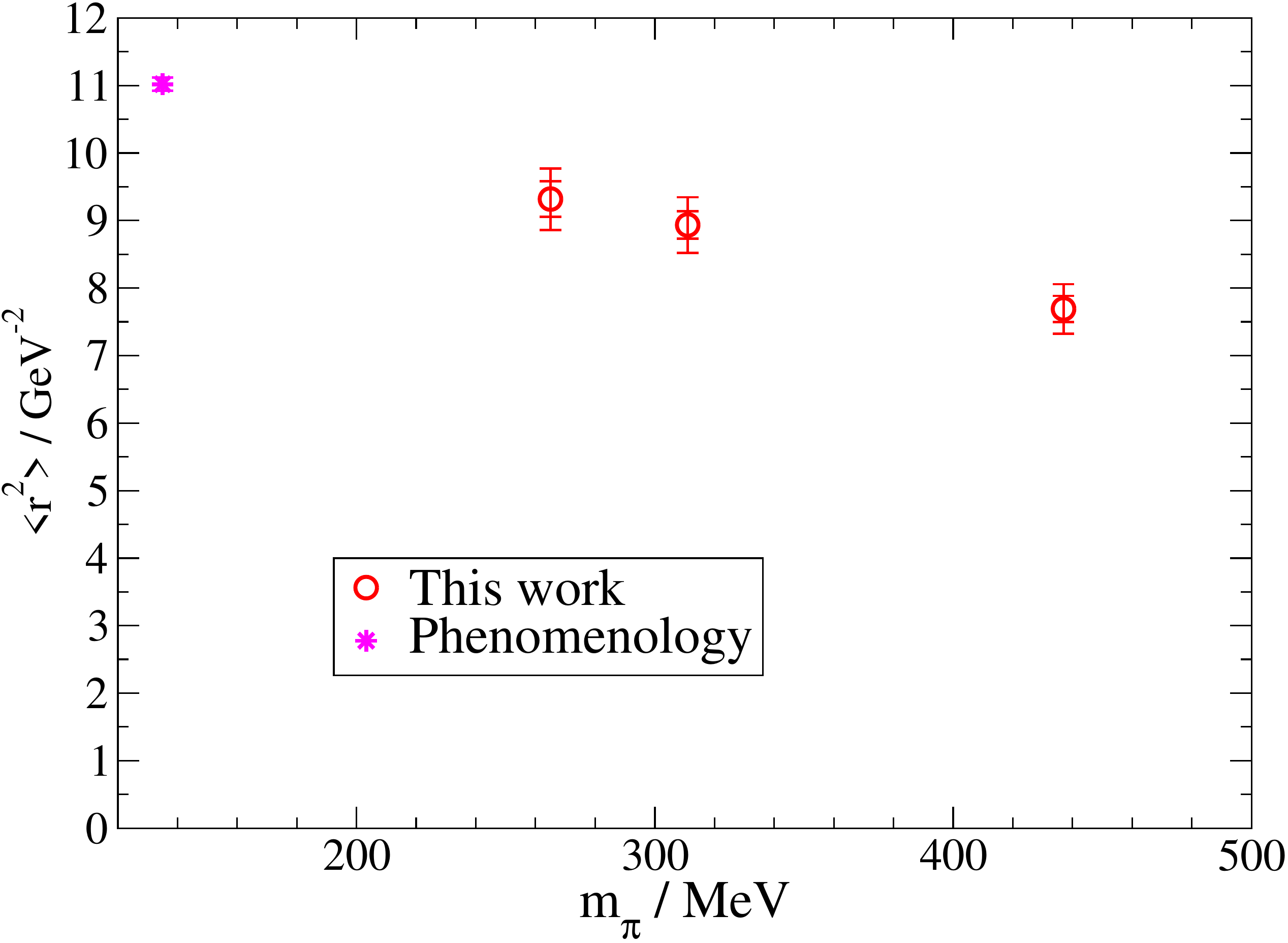}
\includegraphics[width=0.49\linewidth,clip]{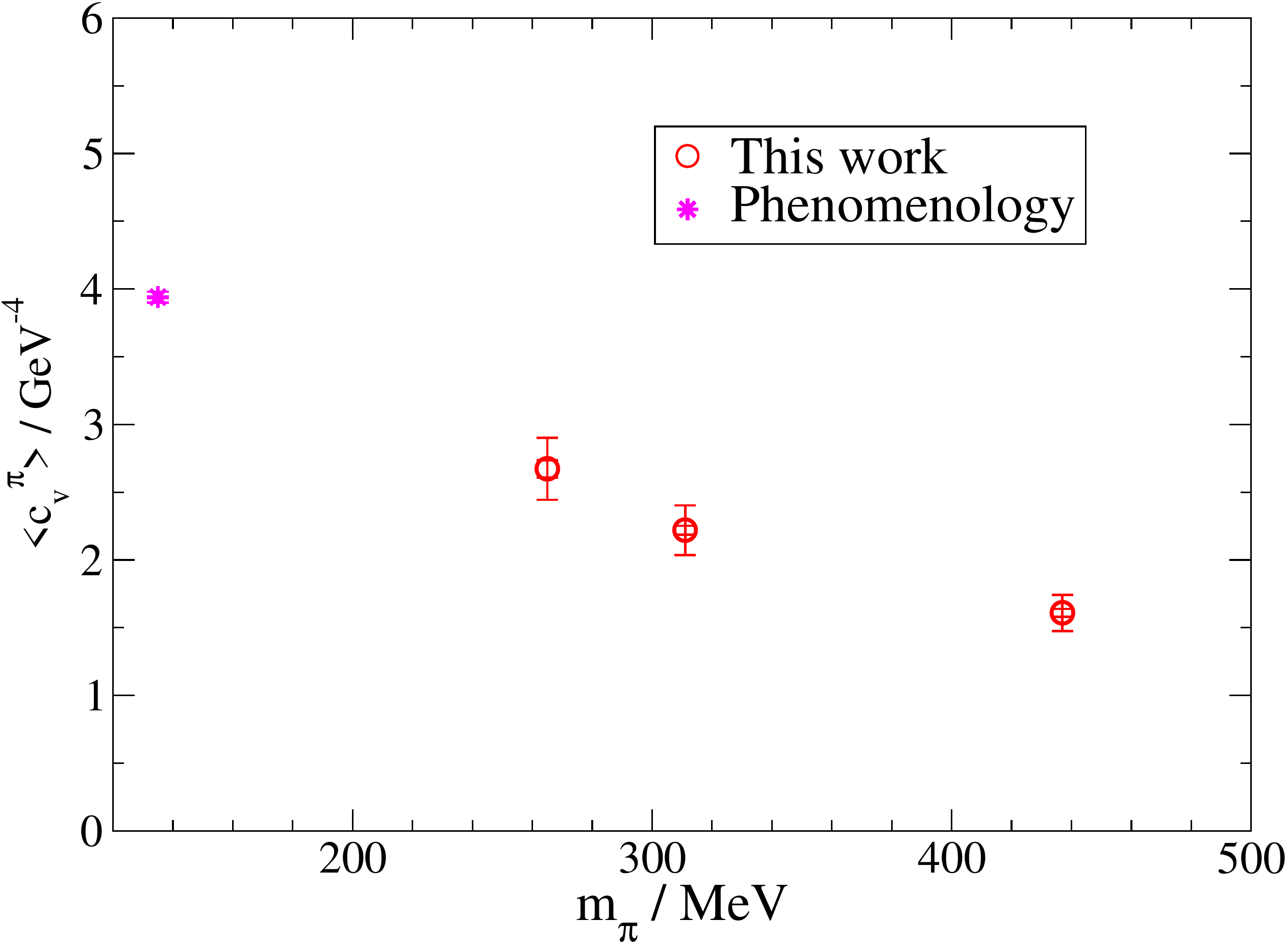}
  \caption{\textbf{Left panel: }Pion mass dependence of the square
    radius $\langle r_\pi^2 \rangle$. \textbf{Right panel: }The same for the
    curvature $c_v$. Our lattice results are compared to the determinations in
    Ref.~\cite{Colangelo:2018mtw} and Ref.~\cite{Gonzalez-Solis:2019iod}
    respectively. The inner error bar on the lattice data denotes the
    statistical uncertainty, while the outer error bar includes the scale
    setting uncertainty from the conversion to physical units.
  }
  \label{figure:radcurv}% Give a unique label
\end{figure} 

As a consistency check our results for the square radius and
  curvature are plotted as a function of the pion mass, along with the values from
  phenomenological determinations in Figure
  \ref{figure:radcurv}. For the square radius we compare to the recent
  determination from Ref.~\cite{Colangelo:2018mtw}, while for the curvature we use the value
  from Ref.~\cite{Gonzalez-Solis:2019iod}, which also provides an overview of various
  determinations. Note that the pion mass dependence of our results is in
  good qualitative agreements with the expectations from
  Ref.~\cite{Guo:2008nc}. Lattice results for the curvature have previously
  been obtained in \cite{Aoki:2009qn}.

\section{Hadronic vacuum polarisation}
\label{hvp}

Recently, it has been realised that the timelike pion form factor has
an important application in the context of lattice calculations of the
hadronic contributions to the muon $g-2$. The hadronic vacuum
polarisation contribution, $\ahvp$, is accessible in lattice QCD via
several integral representations involving the vector correlator
\cite{Meyer:2018til}. A convenient way to evaluate $\ahvp$ is based on
the so-called time-momentum representation
(TMR)~\cite{g-2:tmr-1,g-2:tmr-2,g-2:tmr-3}:
\begin{align}
a_\mu^\mathrm{hvp} = \Big( \frac{\alpha}{\pi} \Big)^2 \int_0^\infty dx_0 \, G(x_0) \tilde{K}(x_0;m_\mu) 
\, ,
\label{eqn:a-mu-integral}
\end{align}
with a known kernel function $\tilde{K}(x_0;m_\mu)$, the muon mass $m_\mu$ and the vector-vector
correlator,
\begin{align}
G(x_0) \delta_{kl} = - \int d^3x \, \langle J^\mathrm{em}_k(x) J^\mathrm{em}_l(0) \rangle
\, .
\end{align}
Here, $J^\mathrm{em}_\mu$ is the electromagnetic current,
\begin{align}
J^\mathrm{em}_\mu(x) = \frac23 \bar u(x) \gamma_\mu u(x) - \frac13 \bar d(x) \gamma_\mu d(x) - \frac13 \bar s(x) \gamma_\mu s(x) + \cdots
\, .
\end{align}
A definition of the kernel can be found in \cite{mainz:g-2}. This correlator can be decomposed into an iso-vector ($I=1$) and an iso-scalar ($I=0$) part, $G(x_0) =
G^{I=1}(x_0) + G^{I=0}(x_0)$. It is also commonly decomposed into connected diagrams from each quark flavour and disconnected diagrams. For comparison with Ref. \cite{mainz:g-2}, we will focus on the connected light-quark contribution, $G^{ud}(x_0) = \frac{10}{9}G^{I=1}(x_0)$.  While for small $x_0$, this correlator can
be precisely computed on the lattice, the signal cannot be traced to
arbitrarily large values of $x_0$, partly due to the deteriorating
signal-to-noise ratio, but also due to the finite time extent of the
lattice. Getting a good estimate for the long-distance behaviour of
$G(x_0)$, which is needed to perform the integral to infinity, is one
of the main challenges. The general idea is therefore to use the
direct lattice data up to some cut-off distance $x_0^\mathrm{cut}$ and to
determine the part above this distance separately.\footnote{One can also obtain rigorous upper and lower bounds for the long-time contribution~\cite{Lehner_RBRC, Borsanyi:2016lpl}, which can be improved with knowledge of the spectral decomposition of $G^{ud}(x_0)$~\cite{Meyer_Lat18, gerardin:2019rua}.} Ref. \cite{mainz:g-2} used a simplistic single-exponential model for the large-time part of $G^{ud}(x_0)$:
\begin{align}
G^{ud}(x_0) = c e^{-m_\rho x_0}
\, ,
\end{align}
where $m_\rho$ was a naive estimate for the rho mass, namely the plateau value of a $\langle\rho(t)\rho^\dagger(0)\rangle$ correlator, and $c$ was determined by fitting $G^{ud}(x_0)$.
We are improving on this method in our work using two different approaches, one using a reconstruction of the finite-volume correlator and one estimating the infinite-volume correlator.

\begin{figure}[htbp] % no figure before 1st section
  \centering
\includegraphics[width=.95\linewidth,clip]{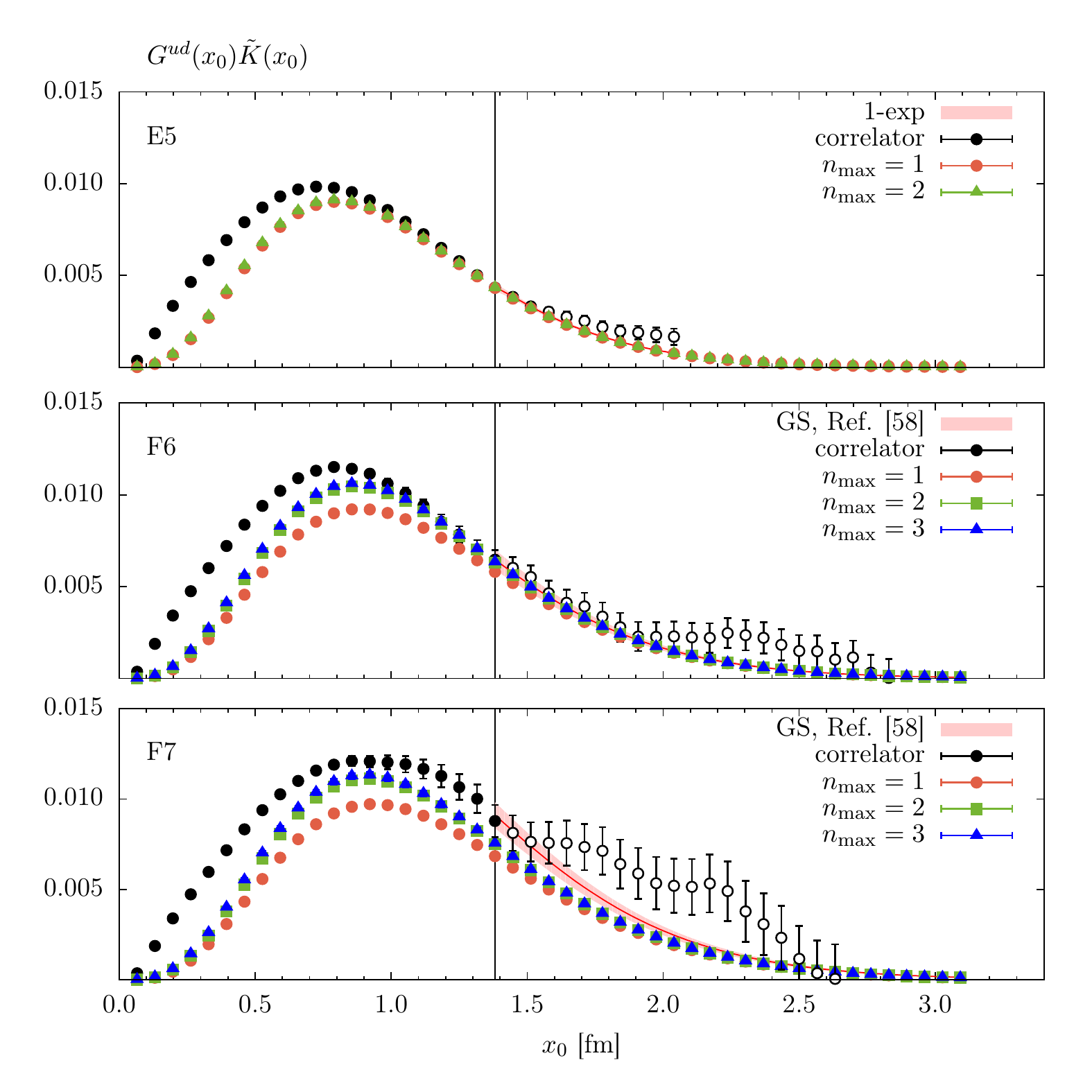}
\vspace{-0.5cm}
  \caption{The light quark contribution to the integrand of Equation
    \eqref{eqn:a-mu-integral} for ensembles E5, F6, F7 (top to
    bottom). The data points computed in \cite{mainz:g-2} are plotted
    as black filled circles up to $x_0^\mathrm{cut}$ and as open
    circles above the cut. The bands represent the continuation of the
    correlator above $x_0^\mathrm{cut}$ as discussed in
    Ref.~\cite{mainz:g-2}. Coloured symbols denote the data from this
    work using the reconstructed light-quark correlator
    $G^{ud}_{n_\mathrm{max}}$ from Equation \eqref{eqn:G-nmax} for
    different values of $n_\mathrm{max}$. The vertical lines indicate
    the value of $x_0^\mathrm{cut}$.}
  \label{figure:a-mu-window}% Give a unique label
\end{figure}

The finite-volume approach uses the information we have about the lowest states in the energy spectrum from the GEVP. We can reconstruct the light-quark correlator with the current matrix elements $|A_{l/c}|$ we already used to compute $F_\pi$,
\begin{align}
G^{ud}_{n_\mathrm{max}}(x_0)= \frac{10}{9}\sum_{n=0}^{n_\mathrm{max}} |A_{lc}|^2_n e^{-E_n x_0} 
\,.
\label{eqn:G-nmax}
\end{align}
This approach has several advantages: Not only do we get a more precise estimate
for the large-$x_0$ behaviour of $G^{ud}(x_0)$, but we also have a way to determine the number of states required for a reliable estimate. By computing
$G^{ud}_{n_\mathrm{max}}(x_0)$ for different values of $n_\mathrm{max}$, we can see
the estimates converging towards each other. In a region where $G^{ud}_{n}(x_0)$
agrees within errors with $G^{ud}_{n+1}(x_0)$, we assume that all energy
levels $n+2$ and above will not contribute significantly to $G^{ud}(x_0)$. The integrand of Equation \eqref{eqn:a-mu-integral} for different values
of $n_\mathrm{max}$ can be seen in Figure \ref{figure:a-mu-window}. We compare
it to the data obtained by a direct calculation of the vector-vector
correlator on the same ensembles, performed in \cite{mainz:g-2}.  Even for
values lower than $x_0^\mathrm{cut}$, the contribution obtained only from the
first level on E5 saturates the contribution from the lowest two levels. On F6
and F7, the contribution from two levels saturates the contribution obtained
from 3 levels, also at comparably low $x_0$. This means that the computation of
further levels would not contribute significantly to $a_\mu^\mathrm{hvp}$, and it also shows that a 1-exponential tail is not well motivated on F6
and F7. Also, on E5 and F6, our reconstructed data saturate the lattice data from \cite{mainz:g-2}
around $x_0^\mathrm{cut}$ and are much more precise afterwards. On F7, the correlator data lie significantly above the reconstruction, which might be caused by
a correlated fluctuation upward that overestimates the vector-vector correlator. Already starting at about $1$ fm, the data from the direct lattice
calculation on F7 seem to deviate from the expected behaviour, leading to
this possible overestimation.

In the infinite-volume approach, the long-time part of the correlator is estimated by evaluating the integral 
\begin{align}
G^{ud}_{F_\pi}(x_0)= \frac{10}{9} \int_{0}^{\infty} d \omega \,
\omega^2 \rho(\omega^2) e^{-\omega x_0} \, .
\label{eqn:vector-vector-corr}
\end{align}
Below the $4m_\pi$ threshold\footnote{Because the integrand is exponentially suppressed at high energy, we use this parameterisation (and the one of $F_\pi$) also above the $4m_\pi$ threshold.}, $\rho(s)$ can be parameterised by
\begin{align}
\rho(s)= \frac{1}{48 \pi^2} \bigg( 1 - \frac{4 m_\pi^2}{s}\bigg)^\frac{3}{2} |F_\pi(s)|^2
\, .
\end{align}
This approach was also used in \cite{mainz:g-2}, where the form factor was estimated using the Gounaris-Sakurai \cite{gounaris-sakurai} parameterisation using the naive rho mass $m_\rho$ and an estimation of the width $\Gamma_\rho$ based on its experimental value and an assumed scaling $\Gamma_\rho \propto k_\rho^3 / m_\rho^2$.\footnote{We will not compare the infinite-volume GS results from Ref. [58] with ours. In that work, the GS model was also used for a finite-volume extension of the correlator, and we compare those results with ours in Table~\ref{table:amu-window}.} 

\begin{figure}[htbp] % no figure before 1st section
  \centering
\includegraphics[width=.95\linewidth,clip]{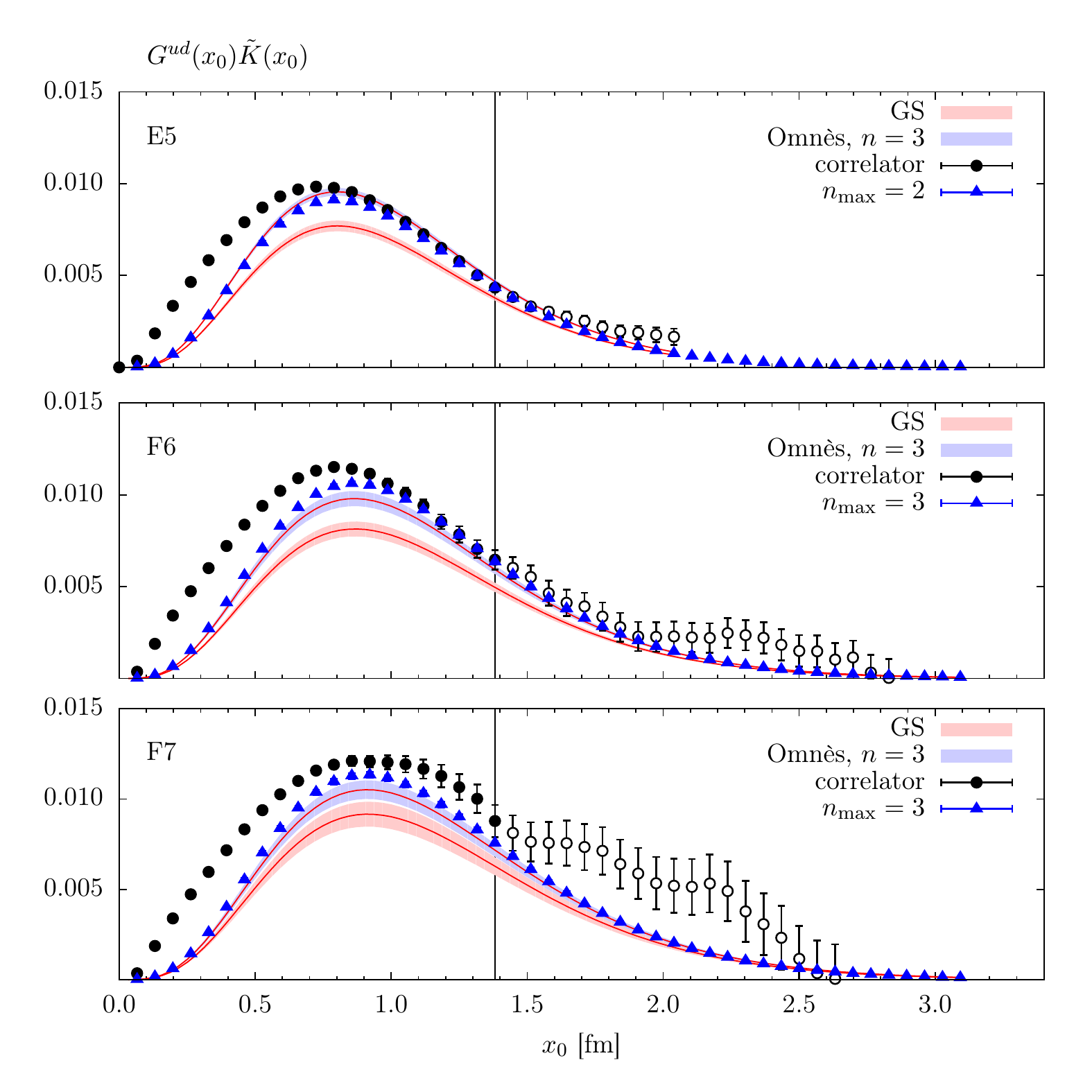}
\vspace{-0.5cm}
  \caption{The integrand of Equation \eqref{eqn:a-mu-integral} for
    ensembles E5, F6, F7 (top to bottom). The meaning of the black
    filled and open circles is the same as in
    Figure~\ref{figure:a-mu-window}. Blue triangles represent the
    integrand reconstructed from the iso-vector correlator
    $G^{ud}_{n_\mathrm{max}}$ of Equation \eqref{eqn:G-nmax} for the
    corresponding value of $n_\mathrm{max}$. Data corresponding to the
    iso-vector correlator constructed from the GS parameterisation and
    the $3$-subtracted Omn\`es representation of $F_\pi$ are shown as
    red and blue bands, respectively. The different types of extending
    the correlator above $x_0^{\mathrm{cut}}$ are used to compute the
    results for $a_\mu^\mathrm{hvp}$ presented in Table
    \ref{table:amu-window}. The difference between the $2$- and
    $3$-subtracted versions of the Omn\`es representation integral is
    too small to be seen on this plot.}
  \label{figure:a-mu-methods-window}% Give a unique label
\end{figure}

In this work, we have several parameterisations of $F_\pi$ and can therefore directly evaluate Equation \ref{eqn:vector-vector-corr}. The result of this is shown in Figure
\ref{figure:a-mu-methods-window}, where we compare the vector-vector
correlator $G^{ud}_{F_\pi}$ obtained from the Gounaris-Sakurai and from the
Omn\`es representation and for comparison show the estimator with the highest $n_\text{max}$ from Figure
\ref{figure:a-mu-window} as well as the Mainz HVP data from \cite{mainz:g-2} again. It is obvious that the Gounaris-Sakurai representation with the resonance parameters from our phase-shift analysis is not a good parameterisation of our data and leads to an integrand that does not saturate the lattice data. 

\begin{figure}[htbp] % no figure before 1st section
  \centering
\includegraphics[width=.8\linewidth,clip]{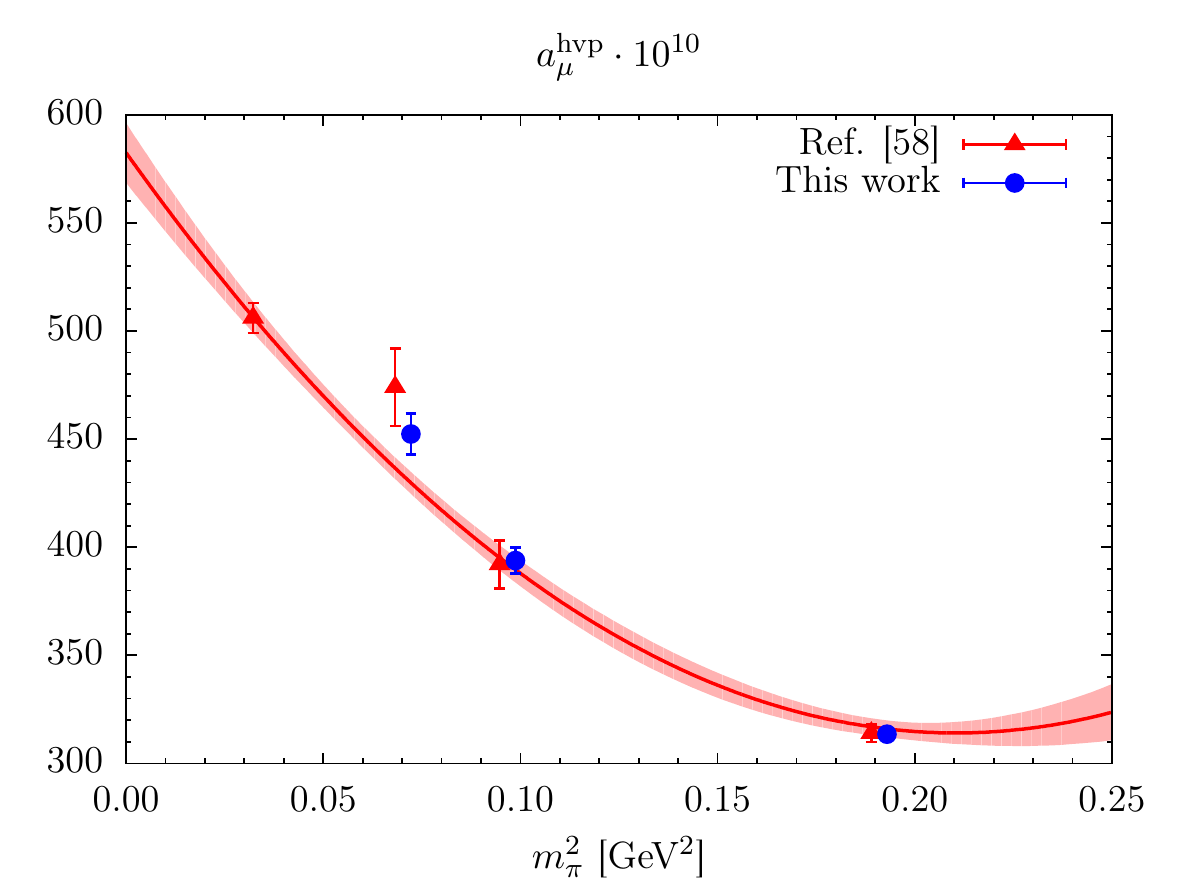}
  \caption{Pion mass dependence of $a_\mu^\mathrm{hvp}$ at
    $\beta=5.3$. Red triangle correspond to the data computed in
    Ref.~\cite{mainz:g-2} with the chiral extrapolation at non-zero
    lattice spacing represented by the band. The blue circles denote
    the data points determined from the large-$x_0$ tail of our most
    precise reconstruction of the correlator, which is the
    $G^{ud}_{n_\mathrm{max}}$ correlator using the matrix elements
    $|A|$ as an input. Points are slightly shifted for clarity.  The
    leftmost red triangle corresponds to ensemble G8 ($m_\pi=185$
    MeV), which was not considered in this work. Our determination of
    the tail of the iso-vector correlator allows for a significantly
    more precise determination of $a_\mu^\mathrm{hvp}$ compared to
    Ref.~\cite{mainz:g-2}. Furthermore, we find that the result for F7
    moves closer to the curve when our reconstruction of the
    large-distance tail is used.}
  \label{figure:amu-chiral}% Give a unique label
\end{figure}

\begin{table}[htbp]
\centering
\begin{tabular}{c|c|c|c}
\hline
\hline
  & E5  & F6  & F7   \\ 
\hline
$0$ to $x_0^\mathrm{cut}$ & 2.662(26) & 3.131(52) & 3.462(86)\\
\hline
$x_0^\mathrm{cut}$ to $\infty$ (1-exp/GS) & 0.484(15) & 0.818(52) & 1.238(96)\\
$x_0^\mathrm{cut}$ to $\infty$ ($G^{ud}_{n_\mathrm{max}}$) & 0.473(9) & 0.808(13) & 1.050(20)\\
$x_0^\mathrm{cut}$ to $\infty$ ($G^{ud}_{F_\pi}, n=2$) & 0.516(13) & 0.776(29) & 1.049(48)\\
$x_0^\mathrm{cut}$ to $\infty$ ($G^{ud}_{F_\pi}, n=3$) & 0.502(13) & 0.805(30) & 1.078(52)\\
\hline
$0$ to $\infty$ (1-exp/GS) & 3.146(39) & 3.949(99) & 4.700(173)\\
$0$ to $\infty$ ($G^{ud}_{n_\mathrm{max}}$) & 3.135(28) & 3.940(59) & 4.524(95)\\
$0$ to $\infty$ ($G^{ud}_{F_\pi}, n=2$) & 3.179(30) & 3.907(63) & 4.511(102)\\
$0$ to $\infty$ ($G^{ud}_{F_\pi}, n=3$) & 3.165(31) & 3.936(65) & 4.540(106)\\
\hline
FV correction, $n=2$ & 0.043(12) &  $-0.032(31)$ &  $-0.001(50)$\\
FV correction, $n=3$ & 0.029(13) &  $-0.003(31)$ &  $ 0.028(54)$\\
FV correction, \cite{mainz:g-2} & & 0.03 & 0.07\\
\hline
\end{tabular}
\caption{Values for $a_\mu^\mathrm{hvp}$ obtained using various methods, in units of $10^{-8}$. The first line shows the accumulated integral over the lattice data up to $x_0^\mathrm{cut}$. The next four lines show the integral over the long-time tail using the following four methods: (1-exp/GS) is the single-exponential (on E5) or the finite-volume GS parametrisation (on F6 and F7), which is a re-analysis of the data from \cite{mainz:g-2}. $G^{ud}_{n_\mathrm{max}}$ is our extension using the reconstruction of the light-quark correlator using Equation \eqref{eqn:G-nmax}. $G^{ud}_{F_\pi}$ reconstructs the vector-vector correlator using Equation \eqref{eqn:vector-vector-corr}, where the pion form factor $F_\pi$ is parametrised by the $n$-subtracted Omn\`es representation for $n=2$ and $n=3$. We do not show the results of $G^{ud}_{F_\pi}$ reconstructed using the GS parameterisation of $F_\pi$ as it does not describe our data well, as can be seen in Figure \ref{figure:a-mu-methods-window}. The last three lines show the estimate of a correction for finite-volume effects, based on the difference between $G^{ud}_{F_\pi}$ and $G^{ud}_{n_\mathrm{max}}$, or based on the GS parametrisations in Ref.~\cite{mainz:g-2} (which was not done for E5).}
\label{table:amu-window}
\end{table}
Table \ref{table:amu-window} shows our results for the long-time tail computed using the different methods employed in this work and compares them to the naive estimate obtained in \cite{mainz:g-2} without access to the resonance data from this work. Also shown is the full value for $a_\mu^\mathrm{hvp}$, which is the sum of the contribution from the direct lattice calculation and the different long-time tails. One can see readily from Figure \ref{figure:a-mu-window} that on F7, our reconstruction of the vector-vector correlator using Equation
\eqref{eqn:G-nmax} does not saturate the data from the direct lattice
computation of $G^{ud}(x_0)$. When comparing our new values for
$a_\mu^\mathrm{hvp}$ with the ones from \cite{mainz:g-2} and the chiral
extrapolation performed in that work (see Figure \ref{figure:amu-chiral}), one
can see that the value for F7 shifts significantly, but that it comes to an
overall better agreement with the chiral extrapolation curve. Because the lattice data on F7 seems to show a large
correlated fluctuation already at about $1$ fm, and because we are using a
transition value of $x_0^\mathrm{cut} \approx 1.38$ fm, the true value for
$a_\mu^\mathrm{hvp}$ might be even lower.  In any case, the published value for F7 lay prominently above the fit curve of the data points sharing the same
lattice spacing and our analysis brought this data point closer to the
curve. A similar issue is observed for the pion radius
when comparing our results in Table \ref{table:r2-window-ll} to the published results in
\cite{mainz:sq-rad}.

Although asymptotically finite-volume effects in $G^{ud}(x_0)$ are
suppressed exponentially as $e^{-m_\pi L}$, in practice these effects
can be significant. When the large-$x_0$ region is dominated by a
small number of states, the volume dependence is not in the asymptotic
regime~\cite{g-2:tmr-1}. Therefore, it is useful to consider the
difference between infinite-volume and finite-volume reconstructions,
which provides an estimate of a finite-volume correction. This is also
shown in Table~\ref{table:amu-window}. On ensemble E5, the correction
is statistically significant and roughly $+1\%$. On F6 and F7, where a
finite-volume correction was previously estimated in
Ref.~\cite{mainz:g-2}, our results are consistent with zero and also
consistent with the previous estimate.

\section{Conclusions}
\label{outlook}
We have performed an analysis of $I=1$ pion-pion scattering on three
different $N_f=2$ ensembles at fixed lattice spacing. Our spectra
have been determined using the variational method for a total of eight
different irreps in the centre-of-mass frame and three different
moving frames with lattice momenta up to $\mathbf{d}^2=3$. The
spectral information was used in a finite-volume analysis to determine
the resonance parameters. We have studied the consistency of different
parameterisations of the phase shift by comparing the results for the
resonance mass $m_\rho$ and the coupling $g_{\rho\pi\pi}$ obtained
from fits to either the Breit-Wigner or the Gounaris-Sakurai
representation. Our results, shown in
Table~\ref{table:phase-shift-window}, indicate that the resonance
parameters are only weakly dependent on the parametrisation.

We have used our parameterisation of the phase-shift together with the
matrix elements of the local and point-split vector currents to
compute the pion form factor, $F_\pi$, in the timelike region. While
the results for $F_\pi$ agree qualitatively with the Gounaris-Sakurai
parameterisation based on our $\rho$ meson masses and couplings, they
are better described by an Omn\`es representation obtained from a fit to
the $F_\pi$ data, taking the resonance parameters $m_\rho$ and
$g_{\rho\pi\pi}$ as input quantities.

The thrice-subtracted version provides a particularly good description
and also allows for the determination of the (squared) pion charge
radius $\langle r_\pi^2 \rangle$. Our results compare well to an
independent calculation of the charge radius on the same ensembles,
obtained from the slope of the pion form factor in terms of the
spacelike momentum transfer $Q^2$ \cite{mainz:sq-rad}. When
  lowering the pion mass, our results for $\langle r_\pi^2 \rangle$
  and the curvature approach the phenomenological values
  \cite{Colangelo:2018mtw,Gonzalez-Solis:2019iod}. The resulting mass
  dependence of the squared radius is compatible with the results in
  Ref.~\cite{Guo:2008nc}.

While the characterisation of resonances using lattice techniques is
interesting in its own right, the gained information can also be put
to good use in different contexts. As another important application we
have considered the calculation of the hadronic vacuum polarisation
contribution to the muon anomalous magnetic moment,
$a_\mu^{\mathrm{hvp}}$. The precision of lattice calculations of
$a_\mu^{\mathrm{hvp}}$ is typically limited by the long-distance tail
of the vector correlator. 

By means of a direct comparison with an earlier study
\cite{mainz:g-2}, we have shown that the precision in
$a_\mu^{\mathrm{hvp}}$ can be substantially increased by describing
the long-distance tail of the TMR integrand (see
Equation~\eqref{eqn:a-mu-integral}) using the spectral information on
the first few states in the iso-vector channel. Alternatively, the
tail of the integrand can be much more accurately constrained via the
representation of the vector correlator in terms of the pion form
factor. These techniques have, in the meantime, been employed in a
recent calculation of $a_\mu^{\mathrm{hvp}}$ on CLS gauge ensembles
with $N_f=2+1$ flavours of dynamical quarks \cite{gerardin:2019rua}.
Going beyond that work, we have used the difference between
infinite-volume and finite-volume reconstructions to estimate
finite-volume effects; the results are consistent with previous
estimates using the Gounaris-Sakurai model.

\acknowledgments
We thank A.Hanlon, B. H\"orz and H. Meyer for useful discussions, and
B. H\"orz for providing a Python interface for TwoHadronsInBox \cite{Morningstar:2017spu}. We are also
grateful to our colleagues within the CLS initiative for sharing ensembles.
Our calculations were partly performed on the high-performance computing cluster Clover at the Institute for Nuclear Physics, University of Mainz and Mogon 2 at Johannes-Gutenberg Universit\"at Mainz.  We thank Dalibor Djukanovic for technical support. The authors gratefully acknowledge the Gauss Centre for Supercomputing e.V. (\url{www.gauss-centre.eu}) for funding this project by providing computing time through the John von Neumann Institute for Computing (NIC) on the GCS Supercomputer JUQUEEN \cite{juqueen} (project HMZ21) at J\"ulich Supercomputing Centre (JSC).

%%%%%%%%%%%%%%%%%%%%%BIBLIOGRAPHY%%%%%%%%%%%%%%%%%%%%%%%%%%%%%%%%
%\bibliographystyle{unsrt}
\bibliography{bib}
\bibliographystyle{apsrev4-1}

\end{document}